\documentclass[aps,prb,showpacs,twocolumn,superscriptaddress]{revtex4-1}
\usepackage{bm,color,amsmath,amssymb,mathrsfs,latexsym,graphicx,psfrag}

\newcommand{\bra}[1]{\left\langle#1\right|}
\newcommand{\ket}[1]{\left|#1\right\rangle}

\newcommand{\bea}{\begin{eqnarray}}
\newcommand{\enea}{\end{eqnarray}}
\newcommand{\beq}{\begin{equation}}
\newcommand{\eneq}{\end{equation}}

\newcommand{\pdag}{\phantom{\dagger}}

\newcommand{\lin}{\notag \\}

\newcommand{\eq}{=&\;}

\newcommand{\bpm}{\begin{pmatrix}}
\newcommand{\epm}{\end{pmatrix}}

\newcommand{\bal}{\begin{align}}
\newcommand{\eal}{\end{align}}

\newcommand{\si}{\;\text{sin}\,}

\newcommand{\co}{\;\text{\text{cos}}\,}

\newcommand{\1}{k_1}

\newcommand{\dg}[1]{#1^{\scriptstyle{\dagger}}}

\newcommand{\sma}[1]{\scriptscriptstyle{#1}}

\newcommand{\GA}{G^{\sma{A}}}
\newcommand{\ano}{{\cal A}_{\sma{U(1)}}}

\newtheorem{theorem}{Theorem}[section]

\newenvironment{proof}[1][Proof]{\begin{trivlist}
\item[\hskip \labelsep {\bfseries #1}]}{\end{trivlist}}

\newcommand{\qed}{\nobreak \ifvmode \relax \else
      \ifdim\lastskip<1.5em \hskip-\lastskip
      \hskip1.5em plus0em minus0.5em \fi \nobreak
      \vrule height0.75em width0.5em depth0.25em\fi}

\begin{document}
\title{Trace Index and Spectral Flow in the Entanglement Spectrum}
 \author{A. Alexandradinata}  
 \affiliation{Department of Physics, Princeton University, Princeton, NJ 08544}
 \author{Taylor L. Hughes}
 \affiliation{Department of Physics, University of Illinois, 1110 West Green St, Urbana IL 61801}
\author{B. Andrei Bernevig} 
 \affiliation{Department of Physics, Princeton University, Princeton, NJ 08544}
 


\begin{abstract}
We investigate the entanglement spectra of topological insulators which manifest edge states on a lattice with spatial boundaries.  In the physical energy spectrum, a subset of the edge states that intersect the Fermi level translates to discontinuities in the trace of the single-particle entanglement spectrum, which we call a `trace index'. We find that any free-fermion topological insulator that exhibits spectral flow has a non-vanishing trace index, which provides us with a new description of topological invariants. In addition, we identify the signatures of spectral flow in the single-particle and many-body entanglement spectrum; in the process we present new methods to extract topological invariants and establish a connection between entanglement and quantum Hall physics. 
\end{abstract}
\date{\today}


\maketitle
The ability to classify a state of matter from knowledge only of  its many-body ground state is an exciting prospect in condensed matter physics. One established method is to evaluate the ground-state expectation values $\Delta_{(i)}\equiv\langle \Psi_{\sma{GS}}\vert M_{(i)}\vert \Psi_{\sma{GS}}\rangle$ for all local operators $M_{(i)}.$ These order parameters distinguish the phase of matter in the Landau symmetry-breaking paradigm.\cite{lubenskybook} Given a set of $\Delta_{(i)}$, we ask: is the phase of matter uniquely determined? We now know it is not, an exceptional case  being fractional quantum Hall states with, for example, filling factors $\nu=1/3$ or $\nu=1/5.$ These states are featureless insulators with all $\Delta_{(i)}$ vanishing, yet they are distinguished by bulk topological properties which can be measured in Hall conductance experiments.\cite{prangebook} 
The discovery that symmetry-breaking does not classify all quantum phases has led to an explosion in the field of topological order.\cite{wenbook} A theory for the full classification of topologically ordered states is not yet understood. In this article we focus on one such classification method: the use of quantum entanglement to classify quantum ground states. 

Quantum entanglement is a recently fashionable approach to classifying myriad states of matter ranging from spin-systems to topological insulators to fractional quantum Hall states.\cite{Hamma2004,levin2006, Kitaev2006,li2008,BrayAli2009,flammia2009, Thomale2010A,Thomale2010B,Pollmann2010,turner2010A,fidkowski2010,Prodan2010,hughes2011, regnault2009, haldane2009,Kargarian2010, lauchli-10prl156404, lauchli-NJP-1367-2630, bergholtz-arXiv1006.3658B,PhysRevLett.106.100405,rodriguez-arXiv1007.5356R,papic-arXiv1008.5087P,hermanns-1009arXiv4199H, PhysRevB.80.201303, 2011arXiv1103.5437Q,2011arXiv1103.0772Z, PhysRevB.83.115322, thomale-2010arXiv1010.4837T, poilblanc-105prl077202, turner2011,  fidkowski-81prb134509, yao-105prl080501, 2010PhRvB..81f4439P, 2008PhRvA..78c2329C, PhysRevB.83.045110,2011arXiv1104.2544S, Poilblanc2011, 2010arXiv1002.2931F, 2011arXiv1104.1139H, Cirac2011, 2011arXiv1105.4808D, PhysRevA.83.013620, 2011arXiv1104.5157D, ryu2006}  The primary classification tools have been the entanglement entropy, the topological entanglement entropy and the entanglement spectrum (ES); these tools are used to reveal subtle non-local correlations in topologically nontrivial systems. 
In this work, we specifically study the  properties of the entanglement spectra of $2D$ time-reversal breaking\cite{haldane1988, thouless1982, qi2006, ProdanJMP2009} and time-reveral invariant\cite{kane2005A,kane2005B,bernevig2006a,bernevig2006c,koenig2007,qi2008B,wu2006,xu2006,fu2007b,moore2007,murakami2004A,roy2009, fu2007a, fu2006, roy2009a,ran2008} topological insulators with spatial entanglement cuts.

If the presence of edges does not break the symmetry that stabilizes the topological insulator (\emph{e.g.} time-reversal, charge-conjugation, point-group symmetries), the physical spectrum must include gapless modes that separate the non-trivial insulator from the vacuum. Our formalism relies on topological insulators that manifest such edge states. There exists topological insulators which cannot be described by this formalism. A case in point is the topological insulator stabilized by inversion symmetry; since inversion symmetry is broken by the presence of edges, such insulators do not exhibit topologically-protected edge states.\cite{hughes2011,turner2011,fu2007a}

If these topologically-protected edge states are present, they interpolate across the energy gap and spectrally connect the conduction and valence bands - this is called  spectral flow. We will show that a subset of edge states at the Fermi level of the physical energy spectrum translates to discontinuities in the trace of the single-particle ES, which we call a `trace index'. We find that any free-fermion topological insulator that exhibits spectral flow has a non-vanishing trace index, which provides us with a new description of topological invariants. In this article our case studies are the Chern insulator and the $Z_2$ time-reversal invariant insulator. We investigate the trace indices in both case studies and demonstrate that the trace index is equivalent to the Chern number ($Z_2$ invariant) for the Chern ($Z_2$) insulator. This formalism extends the work of Ref. \onlinecite{fidkowski2011}, which presents a clear connection between spectral flow in the ES and the bulk topological invariant. 

In addition to the trace index, we identify more signatures of spectral flow in the single-particle and many-body ES; in this process we will establish a connection between entanglement and quantum Hall physics. When an appropriate flux is threaded to induce charge transport, we find that  the change in the full trace of the single-particle ES is exactly the quantity of charge that is pumped. Furthermore, we study insulators with global $U(1)$ symmetry (\emph{i.e.} global charge conservation) so the many-body ES decouples into independent particle-number sectors, forming a `particle-number space'. When charge is pumped in an adiabatic cycle, there is a global translation in the particle-number space by the quantity of pumped charge. In both case studies of the Chern and $Z_2$ insulators, the amount of pumped charge is directly related to the topological invariant. Hence, these signatures in the single-particle and many-body ES separately  provide us with new means to extract the Chern number and the $Z_2$ invariant.


The outline of our paper is as follows. In Sec. \ref{introent} we introduce our chosen geometry, derive the single- and many-body entanglement spectra for free-fermion lattice models and introduce concepts such as an entanglement ground state and an entanglement Fermi level. Our case studies are the Chern insulator  and the $Z_2$ time-reversal invariant insulator (also called the quantum spin Hall insulator). In Section \ref{Chern} we define the $U(1)$ trace index for the Chern insulator and show its equivalence to the Chern number for translationally invariant and disordered systems.  In Section \ref{QSHI} we define the $Z_2$ trace index for the $Z_2$ insulator and show its equivalence to the $Z_2$ invariant. In addition, we identify the signatures of spectral flow in the single- and many-body entanglement spectra of both case studies and establish a connection between quantum Hall physics and entanglement. We make some concluding remarks in Section \ref{conclusion}.

\section{Entanglement spectrum in free-fermion lattice models} \label{introent}

Given a many-body ground state $\ket{\Psi_{\sma{GS}}}$ in a Hilbert space $C$ and its corresponding density operator $\rho = \ket{\Psi_{\sma{GS}}} \bra{\Psi_{\sma{GS}}}$, we partition the Hilbert space into two parts $C = A \otimes B$. The reduced density matrix $\rho_{\sma{A}}$ in the subsystem $A$ may be expressed as the exponential of an entanglement Hamiltonian, ${\cal N} e^{-\hat{H}_{\sma{\text{ent}}}}$, which is normalized such that $\text{Tr}\,\rho_{\sma{A}}=1$. This is likened to the density operator of a thermodynamic system with Hamiltonian $\hat{H}_{\sma{\text{ent}}}$ and unit temperature.\cite{peschel2003}

In this article we distinguish between the \emph{physical} ground state $\ket{\Psi_{\sma{GS}}}$ and the \emph{entanglement} ground state $\ket{\Psi_{\sma{GS}}^{\sma{\text{ent}}}}$, which is the lowest-energy many-body eigenstate of the entanglement Hamiltonian $\hat{H}_{\sma{\text{ent}}}$. The limit of zero entanglement is realized if the many-body ground state $\ket{\Psi_{\sma{GS}}}$ is a single product state in the two subsystems. In this case the entanglement energy gap between $\ket{\Psi_{\sma{GS}}^{\sma{\text{ent}}}}$ and its excited states is infinitely large and the entanglement entropy is zero. A highly entangled $\ket{\Psi_{\sma{GS}}}$ is characterized by small entanglement-energy excitations of $\ket{\Psi_{\sma{GS}}^{\sma{\text{ent}}}}$. 

For a topological insulator, the entanglement entropy of at least one partition $A \otimes B$ cannot be tuned to zero by any change of the physical Hamitonian $H$ that preserves the gap in the physical spectrum and the symmetry that stabilizes the topological phase. This implies that the entanglement energy gap between the entanglement ground state and its excited states cannot be tuned to infinity.\cite{hughes2011} The topology of the phase is thus encoded in the spectrum of $\rho_{\sma{A}}$, which we call the many-body entanglement spectrum (ES).  

For free-fermion Hamiltonians, the \emph{many-body} ES may be derived from the eigenvalues of a reduced one-body correlation matrix \cite{peschel2003,ryu2006}, which we henceforth call the \emph{single-particle} ES. The reduced correlation matrix is defined in Section \ref{sec:singlepart} and its connection to the many-body ES is derived in Section \ref{sect:manybody}. In addition, we discuss the differences between the many-body entanglement spectra of a trivial and topological insulator in Section \ref{sect:manybody}.

\begin{figure}
\centering
\includegraphics[width=8.5cm]{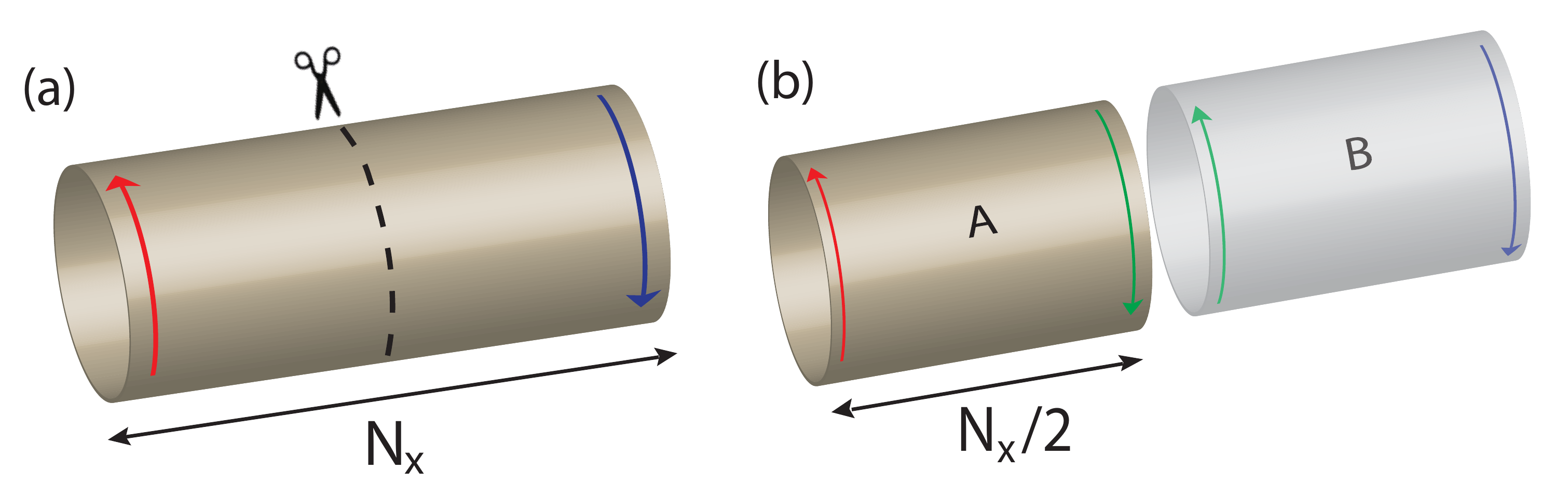}
\caption{(a) Our chosen geometry is a square lattice with open boundary conditions along $\hat{x}$ and periodic boundary conditions along $\hat{y}$: an open-end hollow cylinder. Quantum Hall phases such as the Chern insulator exhibit modes that extend around the edges of the cylinder. Extended states on opposite edges have opposite chirality; the direction of propagation of these modes are indicated by red and blue arrows. (b) We perform a spatial entanglement cut by partitioning the cylinder into two halves - the left half is labelled $A$ and the right $B$. For the Chern insulator, the spectrum of the reduced correlation matrix $G^{\sma{A}}$ in region $A$ includes edge modes (green arrow) on the entanglement cut between $A$ and $B$. These edge modes contribute maximally to the entanglement entropy.}\label{fig:LaughlinCylinder}
\end{figure}

\subsection{Single-particle entanglement spectrum in free-fermion lattice models} \label{sec:singlepart}

 In this section we review the construction of the single-particle entanglement spectrum (ES) from a free-fermion lattice Hamiltonian $H$. We focus on $2D$ lattice models:
\begin{equation} \label{eq:2dham}
H= \sum_{ij,\alpha\beta} h^{\alpha\beta}_{ij}c^{\dagger}_{i\alpha}c^{\pdag}_{j\beta}
\end{equation}\noindent with spatial indices $i,j$ and orbital/spin indices $\alpha,\beta$.  

We will be discussing topological insulators which manifest edge states when placed on a lattice with open boundary conditions. We work with a cylindrical geometry so that the physical spectrum includes states that extend around the edges of the cylinder. We choose $\hat{y}$ to be the azimuthal direction and $\hat{x}$ to lie parallel to the symmetry axis. We have open boundary conditions at sites  $x=1$ and $x=N_x$, as shown in Fig. \ref{fig:LaughlinCylinder}-a. 

If the Hamiltonian is translationally-invariant, we Fourier transform equation (\ref{eq:2dham}) in $\hat{y}$ and obtain
\bal \label{eq:physH}
H = \sum_{k_y,i,j} c_{k_y i \alpha}^\dagger h^{\alpha \beta}_{ij} (k_y) c^{\pdag}_{k_y j \beta}
\end{align}
with $i,j$ now labelling the $\hat{x}$-coordinates of lattice sites. We sum over repeated indices and express the $n$-th single-particle eigenstate of (\ref{eq:physH}) in Bloch form: $\langle y | \psi^n_{k_y} \rangle= e^{ik_yy} [u^n_{k_y}]_{j}^{\alpha} \ket{ j \alpha}$ with $\ket{j \alpha}$  a basis state with orbital/spin $\alpha$ and $\hat{x}$-coordinate $j$. 

The Hamiltonian decouples into irreducible representations labelled by the conserved momentum $k_y$; in each representation, the $n$-th eigenstate of the Bloch Hamiltonian $h^{\alpha \beta}_{ij} (k_y)$ has the corresponding projector $\displaystyle [P^n]_{ij}^{\alpha \beta}(k_y) = [u^{n}_{k_y}]_{i}^{\alpha}  [u^{n\ast}_{k_y}]_{j}^{\beta}$. The many-body ground state wavefunction $\vert\Psi_{\sma{GS}}\rangle$ is a single Slater determinant of all single-particle eigenstates of $H$ with energies less than the Fermi energy $\mu$: $\ket{\Psi_{\sma{GS}}}= \prod_{n, k_y;\varepsilon_n(k_y) < \mu} \dg{\gamma}_{n k_y} \ket{0} $, with normal mode operators $\dg{\gamma}_{n k_y} = [u_{k_y}^{n}]_{j\beta} \dg{c}_{k_y j \beta}$. 

We define a one-body correlation matrix $G$ which is expressible as the complex conjugate of a sum of projectors of occupied single-particle states: 
\bal \label{cormatrix}
[G]_{ij}^{\alpha \beta}(k_y)  \eq \bra{\Psi_{\sma{GS}}} c_{k_y i \alpha}^\dagger c^{\pdag}_{k_y j \beta} \ket{\Psi_{\sma{GS}}} \lin
\eq \sum_{n;\varepsilon_{n}(k_y) <\mu} [P^{n}]_{ij}^{\alpha \beta \ast}(k_y).
\end{align}
Being a sum of projectors, the correlation matrix must itself be a projector, \emph{i.e.} $G(k_y)^2 = G(k_y)$, with eigenvalues $0$ or $1$.

To make a spatial entanglement cut, we partition the cylinder into two halves $A$ and $B$ each of length $N_x/2$; this is illustrated in Fig. \ref{fig:LaughlinCylinder}-b. We choose the convention that region $A$ is the left half of the cylinder. We define the reduced one-body correlation matrix $G^{\sma{A}}(k_y)$ as the correlation matrix $G(k_y)$ with its spatial indices restricted to the subset of sites in region $A$,
\bal \label{reducedcormatrix}
\big[G^{\sma{A}}\big]_{ij}^{\alpha \beta}(k_y)  \eq \bra{\Psi_{\sma{GS}}} c_{k_y i \alpha}^\dagger c^{\pdag}_{k_y j \beta} \ket{\Psi_{\sma{GS}}}; \; i,j \in \big[1,\tfrac{N_x}{2} \big]
\end{align}
If there are $N_{\sma{\text{local}}}$ local degrees of freedom per site, $\GA (k_y)$ has $Q_{\sma{A}} \equiv N_x N_{\sma{\text{local}}}/2$ eigenvalues which we label by $\zeta_i(k_y).$  The spectrum of $G^{\sma{A}}(k_y)$ is bounded by $0$ and $1$. If the Hamiltonian is completely on-site, \emph{i.e.} diagonal in \emph{real} space, the spectrum consists only of $0$'s and $1$'s and consequently the entanglement entropy is zero. This is the atomic limit of an insulator.  We are interested in topological insulators which are defined to be not adiabatically connected to the atomic limit; these phases have finite entanglement entropy which cannot be removed unless the insulator gap collapses or the symmetry that stabilizes the insulator is removed.

Eventually we will describe disordered systems, so we want to define the single-particle ES when there is no translational symmetry in $\hat{y}$. In this case, the correlation matrices $G$ and $G^{\sma{A}}$ do not decouple into different $k_y$ sectors so the generalization is implemented by removing the $k_y$ labels in the definitions in Eq. (\ref{cormatrix}) and (\ref{reducedcormatrix}).

\subsection{Relationship between the many-body and single-particle entanglement spectra of free-fermion models} \label{sect:manybody}

In this section we derive a relation between the many-body entanglement spectrum (ES) and the spectrum of the reduced correlation matrix $G^{\sma{A}}$ and introduce the concepts of an entanglement ground state and an entanglement Fermi energy. We illustrate these ideas with two examples - the entanglement spectra of a trivial and topological insulator - and discuss their differences.

The many-body ES is the set of eigenvalues of the the reduced density operator $\rho_{\sma{A}}$ in region $A$.
If we have translational symmetry in $\hat{y}$, we may factorize $\rho_{\sma{A}}$  into distinct $k_y$ sectors and express it as the exponential of an \emph{entanglement} Hamiltonian ${H}_{\sma{\text{ent}}}$:
\bal
\rho_{\sma{A}} \equiv   \prod_{k_y} \rho_{\sma{A}}(k_y) = {\mathcal{N}} \prod_{k_y} e^{-H_{\sma{\text{ent}}}(k_y)_{ij} \dg{c_{\sma{k_yi}}} c_{\sma{k_yj}}} .
\end{align}
In the rest of the article we let the indices $i,j$ of the matrices $G$,$G^{\sma{A}}$, $H_{\sma{\text{ent}}}$ denote collectively all the degrees of freedom: spatial, spin and/or orbital. 

The matrix $H_{\sma{\text{ent}}}(k_y)$ has $Q_{\sma{A}}$ eigenvalues - the single-particle entanglement energies - which we label by $\xi_{i}(k_y)$; $n_i(k_y)$ are defined as the occupation numbers of the corresponding single-particle eigenstates. ${\mathcal{N}}$ normalizes the reduced density operator such that it has unit trace; in the thermodynamics analogy ${\mathcal{N}}^{-1}$ is the partition function of an entanglement Hamiltonian with unit temperature.
In this analogy, the final form of the reduced density operator assumes a familiar form 
\bal 
\rho_{\sma{A}}(k_y)  \eq \prod_{i=1}^{Q_{\sma{A}}} \frac{e^{-\xi_i(k_y) \dg{\chi_{k_yi}} \chi_{k_yi}}}{1+ e^{-\xi_i(k_y)}}
\end{align}
with normal mode operators $\dg{\chi_{k_yi}} = v^i_j(k_y)  \dg{c_{\sma{k_y j}}}$. 

Let us define $\text{Tr}_{\sma{A},k_y}$ as the trace over all states in subsystem $A$ in the momentum sector $k_y$. We note that $\text{Tr}_{\sma{A},k_y} \rho_{\sma{A}}(k_y)=1$. The reduced correlation matrix as defined in (\ref{reducedcormatrix}) can be expressed with reduced density operators as $G_{ij}^{\sma{A}}(k_y) = \text{Tr}_{\sma{A},k_y}\big[ \rho_{\sma{A}}(k_y) \dg{c_{k_yi}} c^{\pdag}_{k_yj} \big]$. 
We show in Appendix \ref{app:Grho} the final steps to derive 
\bal \label{eigrel}
\big[G^{\sma{A}}(k_y)\big]^{\sma{T}} \eq \frac{1}{1 + e^{H_{\sma{\text{ent}}}(k_y)}},
\end{align}
which relates the single-particle entanglement \emph{energies} ($\xi_i$) to the \emph{eigenvalues} ($\zeta_i$) of $\GA$.
Given a set of occupation numbers $\{n_i(k_y)\}$ for the single-particle entanglement eigenstates, the corresponding eigenvalue $\lambda[\{n_i(k_y)\}]$ of the reduced density operator $\rho_{\sma{A}}$ is
\bal \label{eq:manybody}
& \lambda \bigg[\big\{n_i(k_y)\big\} \bigg] = \prod_{k_y} \prod_{i=1}^{Q_{\sma{A}}} \bigg(1-\zeta_i(k_y) \bigg)\;\bigg(\frac{\zeta_i(k_y)}{1-\zeta_i(k_y)}\bigg)^{n_i(k_y)} \lin
\eq \prod_{k_y} \prod_{i \in \sma{\text{occupied}}} \zeta_i(k_y) \prod_{j \in \sma{\text{unoccupied}}} \big(1-\zeta_j(k_y) \big) 
\end{align}\noindent 
With this relation we can construct the many-body ES (\emph{i.e.} all eigenvalues of $\rho_{\sma{A}}$) by occupying the single-particle entanglement states for each $k_y$ in all possible combinations.

\begin{figure}
\centering
\includegraphics[width=9cm]{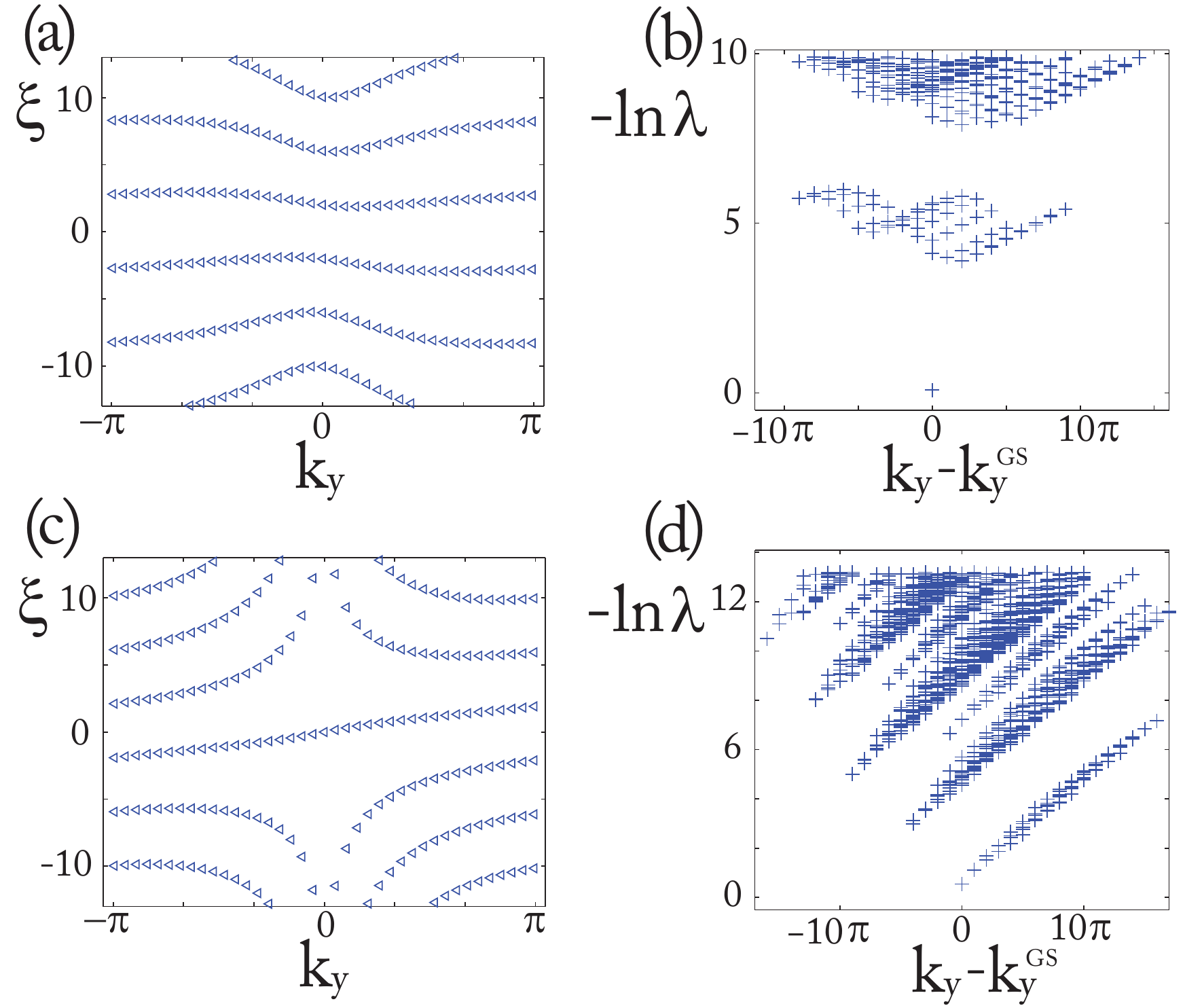}
\caption{ Top: Figures \ref{fig:CImanybodyES}-a and b respectively show the entanglement energy spectrum and the many-body entanglement spectrum (ES) of a trivial insulator that is adiabatically connected to the atomic insulator. The latter is scaled logarithmically. The many-body ES belongs in the half-filled particle number sector, \emph{i.e.} half the total number of single-particle entanglement eigenstates are occupied. Bottom: Figures \ref{fig:CImanybodyES}-c and d respectively show the same spectra for a topological insulator which exhibits spectral flow.  The many-body ES is plotted as a function of the total momenta $k_y$ of the many-body entanglement state. As a reference point, we define $k_{y}^{\sma{GS}}$ as the total momentum of the entanglement ground state.   }\label{fig:CImanybodyES}
\end{figure}

The task of finding structure in the many-body ES is eased by exploiting the global $U(1)$ charge conservation symmetry of the Hamiltonian to decouple the many-body reduced density matrix into different particle number sectors. Let us clarify what this means. The density matrix over the entire Hilbert space is the projector $\ket{\Psi_{\sma{GS}}} \bra{\Psi_{\sma{GS}}}$ of the ground-state many-body wavefunction, which has \emph{one} particle number $N_{\sma{GS}}$ - the number of occupied single-particle states in the ground-state Slater determinant.  When we partition the cylinder we have a finite probability of trapping any number of particles in region $A$; we are only limited by the original number of particles we began with, and how many fermionic particles can fit in $A$. We are often interested in many-body entanglement eigenvalues in a single particle number sector - ${{N}}_{\sma{A}}$.  In this case, we fill up the single-particle entanglement levels using every possible combination of occupation numbers $\{n_i(k_y)\}$ that sum up to $N_{\sma{A}}$ - this is the set of many-body entanglement eigenvalues in the $N_{\sma{A}}$ sector. We define the set of all {nonzero} eigenvalues in the  $N_{\sma{A}}$ sector as
\bal \label{balls}
\bigg\{\lambda \bigg\}_{N_{\sma{A}}} \eq \bigg\{\prod_{k_y} \prod_{i \in \text{occupied}} \zeta_i(k_y) \prod_{j \in \text{unoccupied}} \big(1-\zeta_j(k_y) \big) \lin &\;\;\bigg| \sum_{i,k_y} n_i(k_y) = N_{\sma{A}} \bigg\}.
\end{align}

In analogy with the physical Hamiltonian, we may construct the entanglement ground state in the $N_{\sma{A}}$ sector by filling up  $N_{\sma{A}}$ of the lowest-energy eigenstates of the entanglement Hamiltonian. This ground state has a one-to-one correspondence with the largest many-body eigenvalue $\lambda$ in the $N_{\sma{A}}$ sector through relation (\ref{eq:manybody}). The entanglement Fermi energy $\xi_{\sma{F}}$ is the largest single-particle entanglement energy in the many-body ground state. All other eigenvalues $\lambda$ in the $N_{\sma{A}}$ sector correspond to excited many-body states which derive from particle-hole excitations of the entanglement ground state across this entanglement Fermi level. 

As an example, we plot in Fig. \ref{fig:CImanybodyES}-a the single-particle entanglement energy spectrum $\{\xi\}$ of a trivial translationally-invariant insulator that is adiabatically connected to the atomic insulator. There is a clear spectral gap which separates the lowest positive-$\xi$ band and the highest negative-$\xi$ band. We consider the half-filled entanglement ground state with all the negative-energy single-particle entanglement states filled. We denote the total momenta of all single-particle states in this ground state as $k_y^{\sma{GS}}$.  In Fig. \ref{fig:CImanybodyES}-b we plot the many-body ES $\{\lambda\}$ of the half-filled sector as a function of total momentum $k_y$; more precisely, we have plotted $\{-\text{ln}\,\lambda\}$ with $\lambda$ the eigenvalues of the reduced density matrix in region $A$. Through relations (\ref{eigrel}) and (\ref{eq:manybody}), the entanglement ground state is in one-to-one correspondence with the smallest value of  $-\text{ln}\,\lambda$ in the many-body spectrum at momentum $k_y = k_y^{\sma{GS}}$. All other values of $\lambda$ in the half-filled sector correspond to excited many-body entanglement states which derive from particle-hole excitations across the entanglement gap. Since these particle hole excitations are gapped in the single-particle spectrum of $\{\xi\}$, there is also a clear spectral gap in $\{-\text{ln}\,\lambda\}$ separating the the many-body ground state from its excited states. 

We compare these plots with that of a topological insulator in Fig. \ref{fig:CImanybodyES}-c and d. In the next Section we identify these new plots with a Chern insulator with Chern number $-1$. As shown in Fig. \ref{fig:CImanybodyES}-c, there is a set of single-particle entanglement eigenstates that interpolate across the  gap and connects bands with negative and positive $\xi$. This property is called spectral flow. These single-particle states in the gap extend over the spatial entanglement cut between subsystems $A$ and $B$, so we refer to them collectively as the entanglement edge mode. The entanglement ground state in the half-filled sector has a Fermi level $\xi_{\sma{F}}=0$ which intersects the dispersion of the entanglement edge states. Thus, the low-energy particle-hole excitations of the ground state are gapless and correspond to excitations of the entanglement edge mode. This is reflected in the many-body spectra (Fig. \ref{fig:CImanybodyES}-d) by the absence of a spectral gap between the entanglement ground state and its excited states.  
In the next Section we present our first case study of a topological insulator with spectral flow - the Chern insulator.

\begin{figure*}
\centering
\includegraphics[width=16cm]{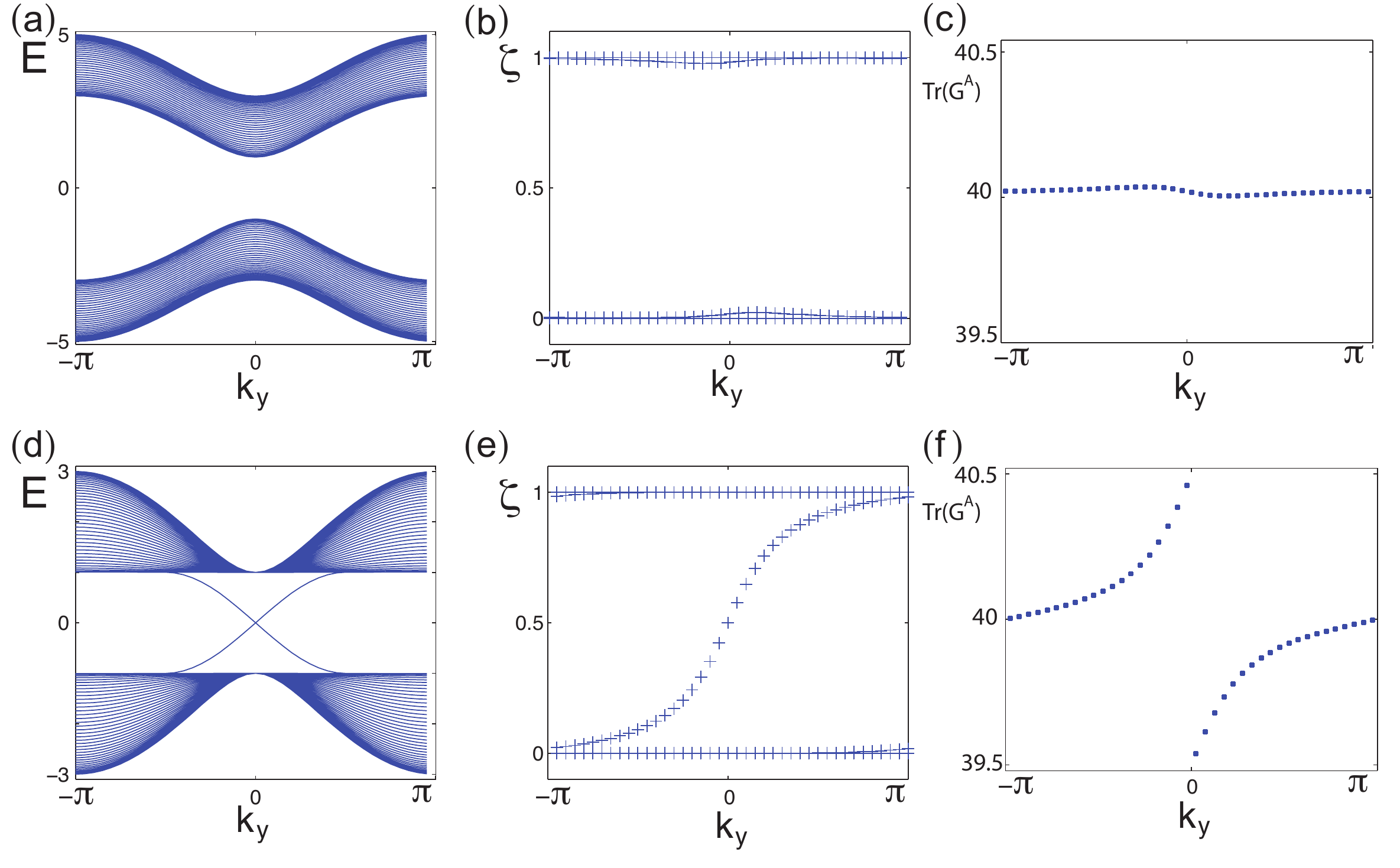}
\caption{Top: Figures 2-a,b,c show, respectively, the physical energy spectrum, the single-particle entanglement spectrum (ES), and the trace of $\GA(k_y)$ for a trivial insulator ($C_1=0$) on an open-end cylinder. In Fig. 2-a and b, bulk bands on opposite sides of the gap are spectrally disconnected from each other. In Fig. 2-c, the trace of $\GA$ is a continuous function of $k_y$ so the trace index ${\cal A}_{\sma{U(1)}}=0$. Bottom: Figures 2-d,e,f show, respectively, the physical energy spectrum, the single-particle ES, and the trace of $\GA(k_y)$ for a nontrivial insulator with $C_1 = -1$. In Fig. 2-d and e, a set of edge states interpolate between bulk bands on opposite sides of the gap - this spectral flow distinguishes trivial and nontrivial Chern insulators. In Fig. 2-f, the trace of $\GA$ drops discontinuously by $1$ at $k_y=0$ - the trace index ${\cal A}_{\sma{U(1)}}=-1$.}\label{fig:CIcomparison}
\end{figure*}

\section{Chern Insulator} \label{Chern}

The Chern insulator (CI) realizes a quantum anomalous Hall effect without application of a net magnetic field.\cite{haldane1988} The simplest CI model exists on a square lattice with two degrees of freedom per site and has the Hamiltonian
\begin{eqnarray} \label{eq:chernham}
H&=&\sum_{(x,y)}\left[\frac{it}{2}\left(c^{\dagger}_{x+1,y}\sigma_1 c^{\pdag}_{x,y}+c^{\dagger}_{x,y+1}\sigma_2 c^{\pdag}_{x,y}-{\text{h.c.}}\right)\right. \lin
&-&\left. \frac{t'}{2}\left(c^{\dagger}_{x+1,y}\sigma_3 c^{\pdag}_{x,y}+c^{\dagger}_{x,y+1}\sigma_3 c^{\pdag}_{x,y}+{\text{h.c.}}\right)\right. \lin
&+&\left.(2-m)\left(c^{\dagger}_{x,y}\sigma_3 c^{\pdag}_{x,y}\right)\right],
\end{eqnarray}\noindent where $c_{x,y}=(c_{x,y\uparrow}\;\; c_{x,y\downarrow})$ destroys a spin up/down fermion on site $(x,y),$ and $t,t',m$ are model parameters. For simplicity we set $t=t'=1$ and the phases of the model are controlled by $m$ alone. Full translational symmetry allows us to express the Hamiltonian as
\bal
H \eq {\displaystyle \sum_{k}} c^{\dagger}_{k}\left[\sin k_x \;\sigma_1 + \sin k_y \;\sigma_2 +M(k)\;\sigma_3 \right]c^{\pdag}_{k} \lin
M(k)\eq 2-m -\cos k_x -\cos k_y.
\end{align}\noindent For $m<0$ and $m>4$ the insulator is trivial and has Chern number $C_1=0.$ For $0<m<2$ the insulator has $C_1=-1$ and for $2<m<4$ the insulator has $C_1=+1.$ To illustrate the different nature of these phases, we plot the physical energy spectra for the $C_1=0$ and $C_1=-1$ cases with our chosen geometry in Figs. \ref{fig:CIcomparison}-a and d respectively. The two spectra are distinguished by gapless chiral edge states which produce a quantum Hall effect in the nontrivial phase.

Let us now examine the single-particle entanglement spectrum of the CI. The eigenvalues of the full correlation matrix $G$ are either  $0$ or $1$ since $G$ is a projector. With an entanglement cut, most of the eigenvalues of the reduced correlation matrix $G^{\sma{A}}$ lie close to $0$ or $1$. For the trivial insulator with $C_1=0$, the bands of eigenvalues near $1$ are spectrally disconnected from the bands near $0$, as shown in Fig. \ref{fig:CIcomparison}-b. If the insulator is nontrivial, there is in addition a set of interpolating single-particle entanglement states that flow across the gap from $0$ to $1$, as shown in Fig. \ref{fig:CIcomparison}-e.  This set of states extend over the entanglement cut between regions $A$ and $B$, and consequently contribute the most to entanglement entropy. The presence of spectral flow is a topological property that guarantees that the entanglement entropy is nonzero for spatial cuts.


\subsection{CI: The trace index and its equivalence to the Chern number} \label{Chernanomaly}

\begin{figure*}
\centering
\includegraphics[width=16.0cm]{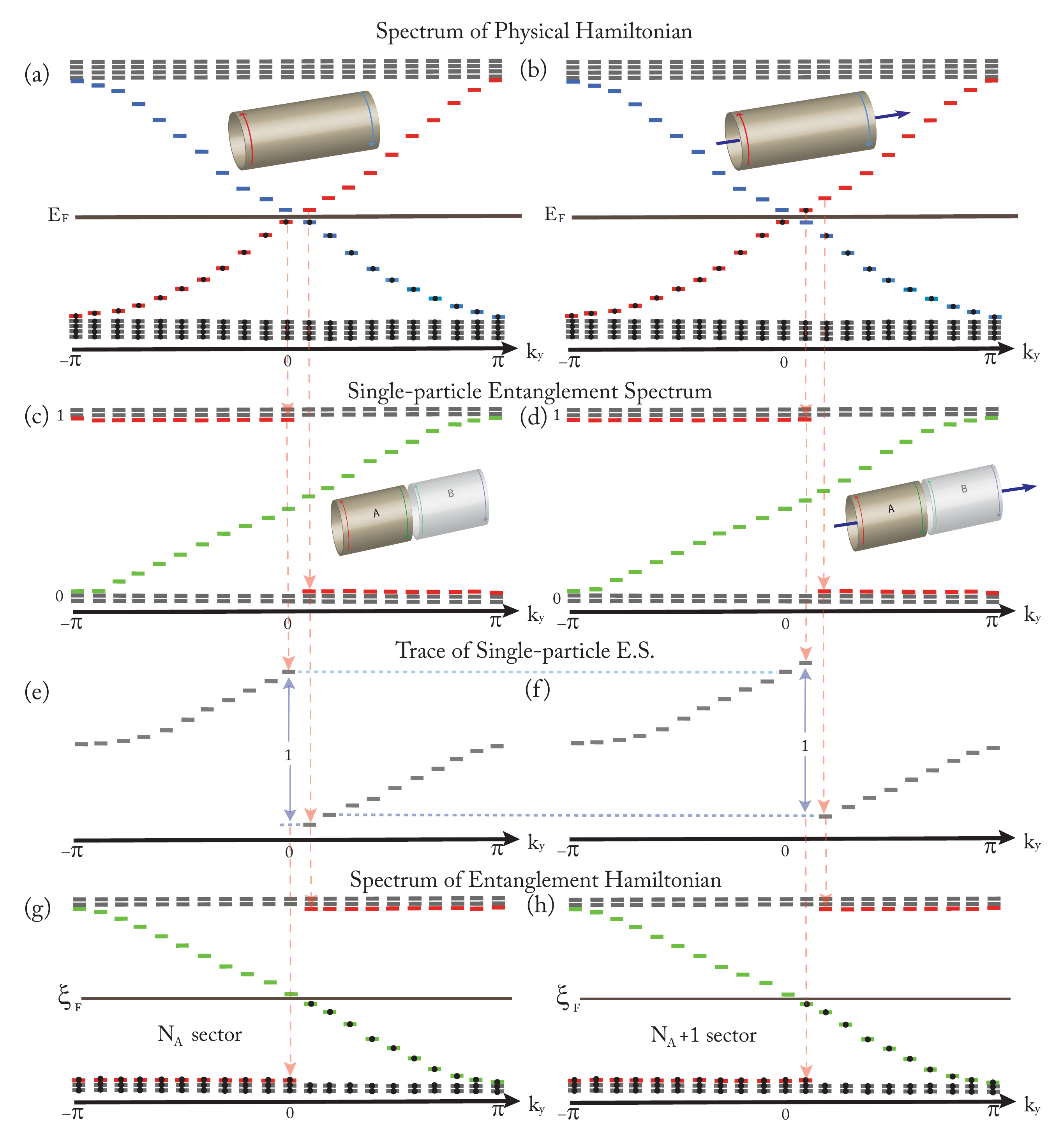}
\caption{Left side: Figures 3-a,c,e,g show schematically the physical energy spectrum, the single-particle entanglement spectrum (ES), the trace of the single-particle ES and the entanglement energy spectrum respectively, as functions of $k_y$, for a many-body ground state of a nontrivial Chern insulator. Right-side: Figures 3-b,d,f,h label the new functions when a flux quantum is inserted through the symmetry axis of the cylinder. These are schematic figures that emphasize importance features in the physical and entanglement energy gaps; bulk bands away from the gaps are conveniently portrayed as dispersionless. Eigenvalues that are colored red, green and blue correspond to eigenstates that extend over the left edge, the entanglement cut and the right edge of the cylinder respectively. Occupied single-particle states of a many-body Slater determinant are labelled by black circles in both the physical and entanglement energy spectra. }\label{CIschematic}
\end{figure*}

In this section we define the trace index for a CI and demonstrate its equivalence to the Chern number. We first discuss a clean, free-fermion insulator.
We illustrate the index for a CI with $C_1 =-1$,  Fermi energy $\mu=0$ and edge states crossing at $k_y=0$ as in Fig. \ref{CIschematic}-a. We consider the trace of the reduced correlation matrix $G^{\sma{A}}(k_y)$:
\begin{eqnarray} \label{numbop}
\text{Tr} \;\big[G^{\sma{A}}(k_y)\big] = \bra{\Psi_{\sma{GS}}} \sum_{i \in A} c_{i,k_y}^\dagger c^{\pdag}_{i, k_y} \ket{\Psi_{\sma{GS}}}.
\end{eqnarray} 
We identify ${\hat{N}}_A(k_y)= \sum_{i\in A} c_{i,k_y}^\dagger c^{\pdag}_{i, k_y}$ as the number operator in region $A$ and in the $k_y$ sector, so we establish that the trace of $G^{\sma{A}}(k_y)$ is the ground-state expectration value of ${\hat{N}}_A(k_y)$. This suggests that the trace is a continuous function of $k_y$ except where modes cross the Fermi level. Since we have assumed we are in a bulk gap, modes crossing the Fermi-level have to be edge states. In this example, one such mode, colored red in Fig. \ref{CIschematic}-a, crosses the Fermi level between $k_y=0$ and $\tfrac{2\pi}{N_y}.$ To the immediate left of the crossing at $k_y=0$, the occupied state just under the Fermi level is on the left edge (up to exponential finite-size corrections) and contributes $+1$ to the trace of $\GA(0)$; to the immediate right at $k_y= \tfrac{2\pi}{N_y}$, the occupied state just under the Fermi level is on the right edge and does not contribute to the trace of $\GA(\tfrac{2\pi}{N_y})$. Hence there is a discontinuity: $\text{Tr}\,\GA(\tfrac{2\pi}{N_y}) - \text{Tr}\,\GA(0) = -1$ which we tentatively identify as the Chern number.

We now consider the most general insulator. The Fermi energy may take on any value in the gap that does not intersect an extended state; this keeps the phase insulating. We define $K_{\sma{\text{cross}}}$ as the average of two adjacent discrete momenta where a left-edge dispersion intersects the Fermi level; in the above example, $K_{\sma{\text{cross}}}=\tfrac{\pi}{N_y}$. If the gradient of the edge dispersion at such an intersection is positive (\emph{i.e.} $\tfrac{\partial \varepsilon}{ \partial k_y}(K_{\sma{\text{cross}}})>0$), there is at least one occupied left-edge state to the left of the crossing at $K_{\sma{\text{cross}}} - \tfrac{\pi}{N_y}$. We claim that this occupied left-edge state is an eigenstate of $G^{\sma{A}}(K_{\sma{\text{cross}}} -\tfrac{\pi}{N_y})$ with eigenvalue exponentially close to $1$, so it contributes $1$ to the trace of $\GA$. We claim that the unoccupied left-edge state just above the Fermi level is an eigenstate of $G^{\sma{A}}(K_{\sma{\text{cross}}} +\tfrac{\pi}{N_y})$ with eigenvalue exponentially close to $0$, so it does not contribute to the trace of $\GA$. As a consequence, the trace of $G^{\sma{A}}$ decreases discontinuously by $1$ across this crossing. An analogous argument leads to the conclusion that a crossing with a negative gradient results in a discontinuous increase by $1$. 

These claims are justified by the following theorem.
\begin{theorem} \label{thm:1}
As defined in equation (\ref{cormatrix}), let $G$ be the one-body correlation matrix of a many-body eigenstate of a free fermionic Hamiltonian $H$. Let the Hilbert space be partitioned into a direct product $A \otimes B$. Let $u^{(L)}$ be a single-particle eigenstate of the Hamiltonian that is confined in $A$. As defined in equation (\ref{reducedcormatrix}), let $G^{\sma{A}}$ be the one-body reduced correlation matrix with indices restricted to $A$. If the many-body wavefunction includes the single-particle state $u^{(L)}$, then $u^{(L)*}$ is an eigenstate of $G^{\sma{A}}$ with eigenvalue $1$. If the many-body wavefunction excludes the single-particle state $u^{(L)}$, then $u^{(L)*}$ is an eigenstate of $G^{\sma{A}}$ with eigenvalue $0$. 
\end{theorem}
The theorem has the following physical motivation: a state that  is localized on the physical edge of $A$ has exponentially vanishing probability to be in region $B$. In the thermodynamic limit, this edge state is fully confined in $A$ and cannot entangle regions $A$ and $B$. As per our discussion in Sec. \ref{sect:manybody}, the single-particle states that do not contribute to entanglement entropy have eigenvalues $0$ or $1$ in the spectrum of $\GA$. The proof of this theorem is presented in Appendix \ref{app:thm1}. 

We apply this theorem to our example of Fig. \ref{CIschematic}-a, where we show that the occupied left-edge state at $k_y=0$ is an eigenstate of $\GA(0)$ with eigenvalue $1$, also colored red in Fig. \ref{CIschematic}-c. The vacant left-edge state at $k_y=\tfrac{2\pi}{N_y}$ in Fig. \ref{CIschematic}-a is shown to be a zero-eigenvalue eigenstate of $\GA(\tfrac{2\pi}{N_y})$ in Fig. \ref{CIschematic}-c. In Fig. \ref{CIschematic}-e we plot the trace of $\GA$ and show that the discontinuity in the trace lies between $k_y=0$ and $ \tfrac{2\pi}{N_y}$ where the left-edge mode intersects the Fermi level.

More generally, there may be $n_{\sma{(+)}}\,(n_{\sma{(-)}})$ left-edge states crossing the Fermi energy at a given crossing $K_{\sma{\text{cross}}}$ with positive (negative) $\tfrac{ \partial \varepsilon}{\partial k_y}$, and each state contributes $-1 (+1)$ to the trace discontinuity. The net discontinuity is $n_{\sma{(-)}} - n_{\sma{(+)}}$. We define the $U(1)$ trace index (${\cal A}_{\sma{U(1)}}$) for clean, free-fermion insulators as the sum of the trace discontinuties over all $K_{\sma{\text{cross}}}$ in the thermodynamic limit: 
\bal \label{sum}
{\cal A}_{\sma{U(1)}} &\equiv\; {\displaystyle \lim_{N_x, N_y \to \infty}} \sum_{K_{\sma{\text{cross}}}} \text{Tr}\;\big[\GA\big] (K_{\sma{\text{cross}}} + \frac{\pi}{N_y}) \lin
&\;\;\;\;- \text{Tr}\;\big[\GA\big] ( K_{\sma{\text{cross}}} - \frac{\pi}{N_y}) \lin
&=\; \sum_{K_{\sma{\text{cross}}}} n_{\sma{(-)}}( K_{\sma{\text{cross}}}) -  n_{\sma{(+)}}( K_{\sma{\text{cross}}})
\end{align}
The index effectively counts the {net} number of {chiral}  single-particle states, on one edge of the cylinder, that intersect the Fermi level. The edge states are responsible for the quantum anomalous Hall effect when flux is inserted through the symmetry axis of the cylinder, or equivalently when an azimuthal electric field is applied. ${\cal A}_{\sma{U(1)}}$ is precisely the number of charges pumped from one edge to another during an adiabatic insertion of a unit flux, \emph{i.e.} $j_i = \epsilon^{ij} \frac{e^2}{h} E_j {\cal A}_{\sma{U(1)}}$. This connection between the Hall current and the trace index is elaborated in Section \ref{section:fluxChern}. In summary, we have identified the $U(1)$ trace index with the first Chern number ($C_1$) of the CI.  
\bal \label{eq:anomalyeqChern}
{\cal A}_{\sma{U(1)}} \eq C_1.
\end{align}
As a confirmation of (\ref{eq:anomalyeqChern}), we numerically plot the trace of $\GA$ for a trivial ($C_1=0$) and nontrivial ($C_1=-1$) insulator respectively in Fig. \ref{fig:CIcomparison}-c and f. The trace is a continuous function of $k_y$ for the trivial case, and drops discontinuously by $1$ at $k_y=0$ for the nontrivial case.

We remark that ${\cal A}_{\sma{U(1)}}$ is quantized to an integer in the thermodynamic limit. The first reason is that physical edge states are perfectly confined in region $A$ in this limit and this is a necessary condition to apply Theorem \ref{thm:1}; for any finite-size lattice the trace index has exponentially small deviations from an integer because an edge state predominantly in $A$ has exponentially vanishing probability to be found in $B$.  The second reason is that physical bulk states do not contribute to the index in this limit. While states in the valence band have finite probability to be in $A$ and thus contribute to the trace of $\GA$, the projectors of the bulk states are a smooth function of $k_y$ in the limit that the discrete spacing $\tfrac{2\pi}{N_y}$ goes to zero. Any discontinuity must arise from a difference in occupation numbers between two adjacent bulk states - this is impossible if the Fermi energy is in the bulk gap.

\subsection{CI: Flux threading and entanglement}

For a nontrivial CI, a time-dependent flux induces an azimuthal electric field which drives a  quantum Hall current. We investigate the effect of flux threading on the single-particle entanglement spectrum in Section \ref{section:fluxChern} and establish a connection between entanglement and quantum Hall physics. In Section \ref{sect:specflowmanybody} we identify the signature of spectral flow in the many-body entanglement spectrum as a global translation by $C_1$ of all particle-number sectors. In Section  \ref{section:fluxChernint} we generalize the trace index to describe disordered CI's by replacing $k_y$ with flux as a pseudo-momentum\cite{niu1985}. In addition we show that a disorder-induced topological phase transition occurs when the disorder strength is increased beyond a critical value. 

\subsubsection{Effect on the single-particle entanglement spectrum} \label{section:fluxChern}

Adiabatically inserting a unit flux $\Phi_0=\tfrac{h}{e}$ into a nontrivial CI leads to an excited many-body state. The flux insertion is implemented by a Peierls's phase which may be gauged away at the cost of twisting the boundary conditions. This twist changes the allowed values of momenta $k_y$ - each single particle state transforms as $\ket{k_y} \rightarrow \ket{k_y + \tfrac{2\pi}{N_y}\frac{\Phi(t)}{\Phi_0}}$ where $\Phi(t)$ is the instantaneous flux at time $t.$ Upon insertion of one flux quantum, the physical energy spectrum is invariant but there is a shuffling of occupation numbers as single-particle states at $k_y$ transform into adjacent states at $k_y + \tfrac{2\pi}{N_y}$. 

We illustrate this process in Fig. \ref{CIschematic}. Before flux-insertion, the Fermi energy lies over a crossing of left- and right-edge mode dispersions at $K_{\sma{\text{cross}}}=\tfrac{\pi}{N_y}$ as shown in Fig. \ref{CIschematic}-a. After flux-insertion, the resultant Slater determinant is shown in Fig.  \ref{CIschematic}-b. Since the hybridization between states on opposite edges is exponentially small, the left-edge state  at $k_y=0$ transforms, upon flux insertion, into a left-edge state at $\tfrac{2\pi}{N_y}$. The right-edge state at $k_y=\tfrac{2\pi}{N_y}$ transforms into a right-edge state at $k_y=\tfrac{4\pi}{N_y}$. These transformations lead to a particle-hole excitation of the ground-state and the resultant wavefunction takes the form $\ket{\Psi_{\Phi_0}}  = \gamma^{\dagger}_L \gamma_R^{\pdag} \ket{\Psi_{\sma{GS}}} $ with $\gamma^{\pdag}_L,\gamma^{\pdag}_R$ normal mode operators of the lowest energy on the left and right edges respectively.  A net charge equal to the Chern number ($-1$ in this example) is pumped from the left to the right edge; this is a consequence of spectral flow in the physical energy spectrum. 

Let $\GA_{\sma{0}}$ and $\GA_{\sma{\Phi_0}}$ denote the reduced correlation matrices before and after insertion respectively; their spectra are plotted in Figs. \ref{CIschematic}-c and d respectively. The difference between the two matrices is that $\GA_{\sma{\Phi_0}}$ has one extra unit eigenvalue and one less zero eigenvalue at $k_y=\tfrac{2\pi}{N_y}$. This difference arises because the flux-inserted state has an extra occupied left-edge state at $k_y=\tfrac{2\pi}{N_y}$. We observe in Fig. \ref{CIschematic}-e that the ground-state satisfies $\text{Tr}\,\GA_{\sma{0}} \big(\tfrac{2\pi}{N_y}\big) - \text{Tr}\,\GA_{\sma{0}} \big(0\big) = -1$, whereas in the flux-evolved state, $\text{Tr}\,\GA_{\sma{\Phi_0}}\big(\tfrac{4\pi}{N_y}\big) - \text{Tr}\,\GA_{\sma{\Phi_0}} \big(\tfrac{2\pi}{N_y}\big) = -1$ as shown in Fig.  \ref{CIschematic}-f. The position of the trace discontinuity at $K_{\sma{\text{cross}}} =\tfrac{\pi}{N_y}$ is translated by $\tfrac{2\pi}{N_y}$ when a unit flux is inserted, but the magnitude of the trace discontinuity and the $U(1)$ trace index ${\cal A}_{\sma{U(1)}}$ is invariant. 

More generally, there may be multiple crossings $K_{\sma{\text{cross}}}$ of the left-edge states with the Fermi level, and each crossing may contribute $\pm 1$ to the trace index depending on the gradients $\tfrac{\partial \varepsilon}{ \partial k_y}(K_{\sma{\text{cross}}})$. When an integer flux is inserted, all the positions of the discontinuities are translated by $K_{\sma{\text{cross}}} \rightarrow K_{\sma{\text{cross}}} + \tfrac{2\pi}{N_y}$ but ${\cal A}_{\sma{U(1)}}$ is invariant. We leave the proof of this claim to Appendix \ref{app:u1inv}. In Section \ref{Chernanomaly} we identified the  ${\cal A}_{\sma{U(1)}}$ with the Chern number of the ground state; in this Section we show that this equality may be extended to any flux-inserted state which {only} differs from the ground state in the occupation numbers of its edge states. This implies that we may {separately} extract the Chern number from both the ground-state and its flux-inserted state through ${\cal A}_{\sma{U(1)}}$.


We now present an {equivalent} way of extracting the Chern number from two many-body states differing in flux. While the physics is identical, we focus on a different quantity in the entanglement spectrum to emphasize the connection with the quantum Hall effect. This quantity is the \emph{full} trace of $G^{\sma{A}}$ over all $k_y$ sectors $\sum_{k_y} \text{Tr} \;G^{\sma{A}}(k_y) = \bra{\Psi} \sum_{i \in A, k_y} c_{i, k_y}^\dagger c^{\pdag}_{i, k_y} \ket{\Psi} = \bra{\Psi}{\hat{N}}_{A} \ket{\Psi}$. This is the expectation value of the total number of particles in region $A$. We claim that
\beq \label{QHcurrent}
\sum_{k_y} \text{Tr}\;\left[\GA_{\sma{\Phi_0}}(k_y)\right]  - \sum_{k_y} \text{Tr}\;\left[ \GA_{\sma{0}} (k_y)\right] = -C_1. 
\eneq
Proof: under flux insertion, the positions of the discontinuities in the trace of $\GA(k_y)$ are translated by $\tfrac{2\pi}{N_y}$ but the magnitudes of the discontinuities are invariant. This implies that each discontinuity of $-1\,(+1)$  when translated forward adds (deducts) $\tfrac{2\pi}{N_y}$ to the area under the piecewise-continuous function $\GA(k_y)$. The net decrease in the area under $\GA(k_y)$ is the sum of the discontinuities. The full trace of $\GA$ is just the area under $\GA(k_y)$ in units of $\tfrac{2\pi}{N_y}$. By definition, the sum of the discontinuities is the index ${\cal A}_{\sma{U(1)}}$ which is also the Chern number. 

We remark that Tr$\,\big[\GA \big]$ and the charge pumped from $B$ to $A$ during an adiabatic cycle are well-defined quantities independent of the details of the Hamiltonian. In disordered and/or interacting systems, $k_y$ is not a good quantum number but $\text{Tr}\;\left[\GA_{\sma{\Phi_0}}\right]  -  \text{Tr}\;\left[ \GA_{\sma{0}}\right] = -C_1$ still applies. We substantiate by reproducing (\ref{QHcurrent}) through a quantum Hall response. The time-dependent flux  generates an azimuthal electric field 
\bal
E_y \eq - \frac{\partial A_y}{\partial t} = -\frac{1}{ \,N_y} \frac{\partial \Phi}{\partial t}.
\end{align} 
For any CI (interacting/free, disordered/clean), a quantum Hall current flows in response to the field $E_y$: 
\bal
j_x \eq \sigma_{\sma{H}} E_y = \frac{e}{N_y} \frac{\partial \; \bra{\Psi}\, \hat{N}_{\sma{A}}\, \ket{\Psi}}{ \partial t} = \frac{e}{N_y} \frac{\partial\; \text{Tr}\;\big[\GA \big]}{ \partial t}.
\end{align} 
This current is shown to be proportional to the rate of change of the full trace of $\GA$. This implies that the change in flux $\Phi$ is proportional to the change in Tr$\,\big[\GA \big]$. When we insert one flux quantum $\tfrac{h}{e}$, the change in Tr$\,\big[\GA \big]$ equals $- \tfrac{h}{e^2} \sigma_{\sma{H}} = -C_1$.

\subsubsection{Spectral flow and the many-body entanglement spectrum} \label{sect:specflowmanybody}

How spectral flow is manifested in the physical energy spectrum is well-understood - the global spectrum is invariant under insertion of a unit flux but $C_1$ charge is pumped resulting in a quantum Hall current. In this section we investigate how spectral flow manifests itself in the many-body entanglement spectrum (ES). We study insulators with global $U(1)$ charge conservation symmetry so the many-body ES decouples into independent particle-number sectors, forming a `particle-number space'. We find that, analogously, the many-body ES is invariant under insertion of a unit flux and the charge pumping is manifested in a global translation of $C_1$ units in particle-number space.  

We choose the convention that $C_1$ charge is pumped from region $A$ to $B$ when a unit flux is inserted, so there are $C_1$ less occupied left-edge states in region $A$. Theorem \ref{thm:1} informs us that in the thermodynamic limit the single-particle ES now has $C_1$ less unit eigenvalues and $C_1$ more zero eigenvalues. Let us illustrate this with the same example of a $C_1=-1$ insulator with a Fermi energy ($\mu=0$) that intersects the left-edge mode dispersion at $K_{\sma{\text{cross}}}=\tfrac{\pi}{N_y}$. For this example, the spectra of the entanglement Hamiltonians before and after flux insertion are plotted in Figs. \ref{CIschematic}-g and h respectively. The spectra in Figs. \ref{CIschematic}-g and h are respectively related to the spectra of $\GA$ in Figs. \ref{CIschematic}-c and d through the relation (\ref{eigrel}). We first consider the $N_{\sma{A}}$-number sector of the many-body ES before flux insertion. As illustrated in Fig. \ref{CIschematic}-g, the entanglement ground state of this sector is constructed by filling up $N_{\sma{A}}$ of the lowest-energy eigenstates of the entanglement Hamiltonian; we call the largest energy of the occupied states the entanglement Fermi energy $\xi_{\sma{F}}$. The many-body entanglement spectra in this number sector is derived by applying Eq. (\ref{balls}) to the entanglement ground-state and all possible particle-hole excitations across the entanglement Fermi level. We denote the set of nonzero eigenvalues in this sector by $\{ \lambda(\Phi=0) \}_{N_{\sma{A}};\xi_{\sma{F}}}$. 

When a unit flux is threaded, a right-edge state is transported to the left edge. The difference between Fig. \ref{CIschematic}-c (before) and d (after) is that a $0$ eigenvalue of $\GA$ at $\tfrac{2\pi}{N_y}$ is raised across the entanglement gap and now has value $1$. Equation (\ref{eigrel}) informs us that $\zeta(\tfrac{2\pi}{N_y})=1$ corresponds to an entanglement energy $\xi(\tfrac{2\pi}{N_y}) = -\infty$. In Fig. \ref{CIschematic}-h (after) we show that the entanglement ground state formed from occupying all single-particle states under the \emph{same} entanglement Fermi level  $\xi_{\sma{F}}$  now has an additional particle with $\xi = -\infty$ at $k_y = \tfrac{2\pi}{N_y}$.

We ask what is the spectrum obtained from particle-hole excitations across the \emph{same} Fermi level $\xi_{\sma{F}}$ but now with $N_{\sma{A}}+1$ particles? In other words, what is $\{ \lambda(\Phi = \Phi_0) \}_{N_{\sma{A}}+1;\xi_{\sma{F}}}$? We claim that it is identical with $\{ \lambda(\Phi=0) \}_{N_{\sma{A}};\xi_{\sma{F}}}$ in the thermodynamic limit. 

The physical argument is as follows: we have shown that single-particle eigenstates of the physical Hamiltonian that are confined in $A$ cannot entangle regions $A$ and $B$, thus they have eigenvalues $\zeta_i=0$ or $1$. Having more or less of these zero-entanglement states as a result of flux insertion cannot affect the entanglement entropy. Since the two sets of nonzero eigenvalues, $\{ \lambda(\Phi=0) \}_{N_{\sma{A}};\xi_{\sma{F}}}$ and $\{ \lambda(\Phi = \Phi_0) \}_{N_{\sma{A}}+1;\xi_{\sma{F}}}$, differ only by one zero-entanglement single-particle state at $\tfrac{2\pi}{k_y}$, the two sets must be identical for the entanglement entropy to be invariant. 
 Let us define the many-body wavefunctions before and after insertion of the unit flux as  $\ket{\Psi_{\sma{0}}}$ and $\ket{\Psi_{\sma{\Phi_0}}}$ respectively. The number sectors before and after insertion are related by 
\bal \label{specflowentspec}
\bigg\{\lambda \bigg\}_{N_{\sma{A}}}^{\ket{\Psi_{0}}} \eq \bigg\{\lambda \bigg\}_{N_{\sma{A}}-C_1}^{\ket{\Psi_{\sma{\Phi_0}}}}. 
\end{align} The mathematical proof is presented in Appendix \ref{app:n0n1}.
This is a global translation by $C_1$ in particle-number space, i.e. all particle-number sectors $N_{\sma{A}}$ flow to $N_{\sma{A}}-C_1$, which is analogous to the charge pumping in the physical energy spectrum. This is a change in the labels of each sector; the eigenvalues in each sector are invariant. As a consequence, the entanglement entropy, which depends only on the values of $\lambda_i$, is invariant under integer flux insertion.


The derivation of (\ref{specflowentspec}) through Theorem \ref{thm:1} and the proof in Appendix \ref{app:n0n1} is applicable to any free-fermion Hamiltonian, so the particle-number flow occurs for disordered Chern insulators as well. If we introduce interactions, this derivation no longer applies but the principle remains: with regard to the many-body entanglement eigenvalues, the many-body ground state with $C_1$ less/more particles on the edge of $A$ is indistinguishable from the many-body ground state before flux insertion, because the difference in the two wavefunctions lies far away from the entanglement cut between $A$ and $B$. If the interactions preserve the global $U(1)$ symmetry, the many-body ES still decouples into particle-number sectors which flow as in (\ref{specflowentspec}). 
We also remark that equation (\ref{specflowentspec}) is derived for the integer quantum Hall effect. However, spectral flow is generic in all quantum Hall states and we have an analogous result for the fractional quantum Hall states if we insert more than one unit flux in analogy with the charge pumping picture presented in Ref. \onlinecite{tao1984}.

\subsubsection{The trace index of disordered systems and a disorder-induced topological phase transition}  \label{section:fluxChernint}

The effect of disorder on topological insulators is an ongoing field of resarch.\cite{Obuse2008,Nomura2007,Ryu2007,essin2007,Onoda2007,Sheng2006, Prodan2010} 
Generic topological insulators are characterized by extended bulk states that resist Anderson localization; these insulators are robust against disorder provided some symmetries are preserved.\cite{schnyder2008A}  For the nontrivial CI on a cylinder, we introduce disorder by adding a random spin-dependent potential on each site. The disordered Hamiltonian takes the form
\bal
H_{\omega} \eq H_o + W \sum_{x,y,\alpha} \omega_{x,y,\alpha} \;\dg{c}_{x,y,\alpha} \;c^{\pdag}_{x,y,\alpha}
\end{align} 
with $H_o$ the lattice Hamiltonian introduced in Eq. (\ref{eq:chernham}). $\alpha$ denotes a spin index and $w_{x,y,\alpha}$ is a random number between $-0.5$ and $+0.5$. We tune the strength of this disorder potential by a parameter $W$. 

The trace index $\ano$ as defined in (\ref{sum}) for clean, free systems relies on the conserved quantity $k_y$. However, we have shown in (\ref{sum}) that $\ano$ effectively counts the net number of chiral single-particle states, extended over either edge of the cylinder, that intersect the Fermi level. To generalize the description of trace indices to disordered free-fermion topological insulators, we replace $k_y$ with an adiabatic parameter - the flux inserted through the symmetry axis of the cylinder - that acts as a pseudo-momentum.

The following discussion mirrors the introduction of $\ano$ in Section \ref{Chernanomaly} and is conceptually similar. In analogy with the Fermi energy of a clean insulator, we choose an energy $\varepsilon_{\sma{F}}^{\sma{\Phi}}$ in the mobility gap of the disordered energy spectrum. For each value of the flux $\Phi$, we define the many-body wavefunction $\ket{\Phi}$ as the Slater determinant of all single-particle states with energies less than $\varepsilon_{\sma{F}}^{\sma{\Phi}}$: $\ket{\Phi} = \prod_{n;\varepsilon_n(\Phi) < \varepsilon_{\sma{F}}^{\sma{\Phi}}} \dg{\gamma}_{n,\Phi} \ket{0}$. If the insulator exhibits spectral flow, inserting integer flux into some initial ground state $\ket{\Phi=0}$ creates particle-hole excitations. In this case, $\ket{\Phi}$, which is defined as a fully-filled Fermi sea for all values of the flux, is not necessarily the flux-evolved state that is adiabatically connected to $\ket{\Phi=0}$, \emph{i.e.} we are specifying for $\ket{\Phi}$ that all of the lowest-energy single-particle states are filled regardless of how they adiabatically evolve. The reduced correlation matrix defined as
\bal
\big[C^{\sma{A}}\big]_{ij}^{\alpha \beta}(\Phi)  \eq \bra{\Phi} c_{i \alpha}^\dagger c^{\pdag}_{j \beta} \ket{\Phi}; \;\; i,j \in A
\end{align}
is analogous to but different from our previous definition of $\GA$ in (\ref{reducedcormatrix}). $i,j$ are lattice sites with the $\hat{x}$-coordinate restricted to lie in the range $[1,N_x/2]$.

We define the $U(1)$ trace index ($\ano^{\sma{\Phi}}$) for disordered insulators as the sum of the discontinuities in the trace of $C^{\sma{A}}$ as a function of the flux $\Phi$ in the range $[0,2\pi)$. These discontinuities arise when states in region $A$ disperse with $\Phi$ and intersect $\varepsilon_{\sma{F}}^{\sma{\Phi}}$. There exist localized states in the mobility gap which are confined in $A$ but do not extend around the circumference of the cylinder. The effect of flux on such a state is a change of gauge: its wavefunction acquires a phase but its energy is invariant. These localized states do not disperse as a function of $\Phi$ and cannot intersect $\varepsilon_{\sma{F}}^{\sma{\Phi}}$. If the insulator is topological and exhibits spectral flow, there exist phase-coherent states that extend around the physical edge of $A$. For such a state, the effect of a non-integer flux cannot be gauged away, lest the phase of the wavefunction become discontinuous at a point along the edge of $A$. The energies of these edge states change as a function of $\Phi$ and may intersect $\varepsilon_{\sma{F}}^{\sma{\Phi}}$. Let $\Phi_{\sma{\text{cross}}}$ label the values of flux where a  single-particle state on the edge of $A$ intersects $\varepsilon_{\sma{F}}^{\sma{\Phi}}$ in the range $\Phi \in [0,2\pi)$. Let $n_{\sma{(+)}}( \Phi_{\sma{\text{cross}}})$ ($n_{\sma{(-)}}( \Phi_{\sma{\text{cross}}})$) be the number of left-edge states that cross $\varepsilon_{\sma{F}}^{\sma{\Phi}}$ with positive (negative) $\tfrac{\partial \varepsilon}{\partial \Phi}$ at $\Phi_{\sma{\text{cross}}}$. Then from our definition of the disordered trace index,
 \bal
 \ano^{\sma{\Phi}} &\equiv \; {\displaystyle \text{lim}_{\delta \rightarrow 0^{+}}} \sum_{\Phi_{\sma{\text{cross}}}} \text{Tr}\;\big[C^{\sma{A}}\big] (\Phi_{\sma{\text{cross}}} + \delta) \lin
&\;\;\;\;- \text{Tr}\;\big[C^{\sma{A}}\big] ( \Phi_{\sma{\text{cross}}} - \delta) \lin
&=\; \sum_{\Phi_{\sma{\text{cross}}}} n_{\sma{(-)}}( \Phi_{\sma{\text{cross}}}) -  n_{\sma{(+)}}( \Phi_{\sma{\text{cross}}})
\end{align}
 The index counts the net number of chiral single-particle states that extend around the physical edge of $A$ and whose energies intersect $\varepsilon_{\sma{F}}^{\sma{\Phi}}$. This number is independent of the value of $\varepsilon_{\sma{F}}^{\sma{\Phi}}$ as long as $\varepsilon_{\sma{F}}^{\sma{\Phi}}$ remains in the mobility gap.   The presence of these edge states at $\varepsilon_{\sma{F}}^{\sma{\Phi}}$ indicate a quantum Hall response and we identify $\ano^{\sma{\Phi}}$ with the Chern number.  

\begin{figure}
\centering
\includegraphics[width=8.5cm]{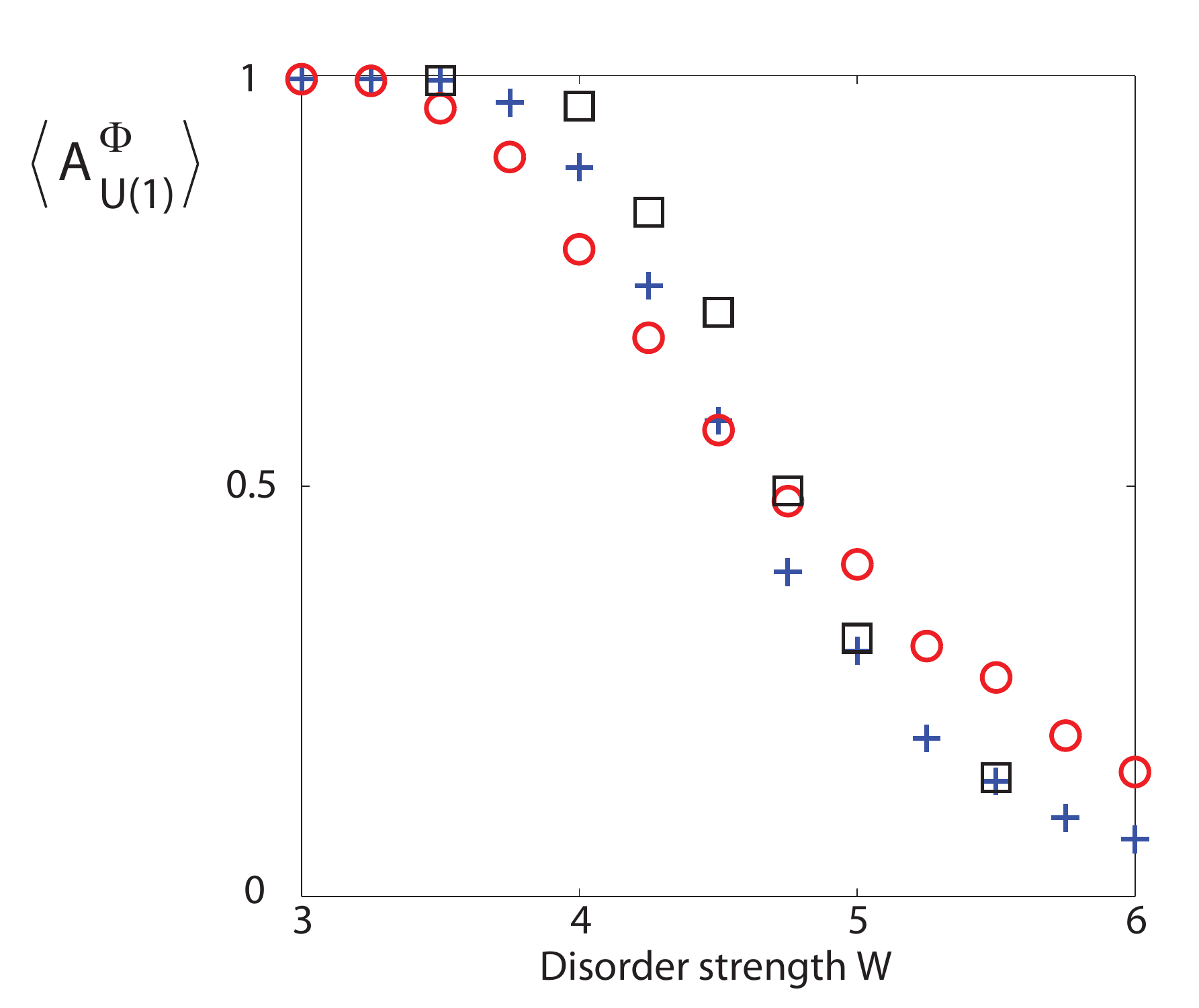}
\caption{Disorder-quenched trace index for the Chern insulator as a function of the disorder strength parameter $W$. The indices are plotted for cylinders of dimensions: $N_x=10$ by $N_y=5$ (circles), $25 \times 25$ (crosses) and $40 \times 40$ (squares). We observe an increasingly sharp transition from the nontrivial insulator with (with trace index $1$) to the trivial insulator (with trace index $0$) as the dimension of the cylinder is increased.}\label{fig:disorderCI}
\end{figure}

If we begin with a nontrivial Chern insulator, we demonstrate that this insulator undergoes a disorder-induced phase transition to a trivial insulator as we increase the disorder strength $W$. We average the trace index of the disordered CI over $400$ random configuration of $\omega_{x,y,\alpha}$ with fixed $\varepsilon_{\sma{F}}^{\sma{\Phi}}$. The disorder-quenced trace index is plotted as a function of $W$ in Fig. \ref{fig:disorderCI} for cylinders with dimensions $N_x=10$ by $N_y=5$, $25 \times 25$ and $40 \times 40$. As the dimensions of the cylinder are increased, we observe an increasingly sharp phase transition from the nontrivial insulator (with trace index $1$) to the trivial insulator (with trace index $0$); this occurs at around $W_c = 5$ in units $t=t'=1$.

The nonzero trace index up to the critical value of $W_c$ demonstrates that the disordered CI has extended edge states that resist Anderson localization for moderately strong disorder. There remains extended bulk states that are spectrally connected to the edge states which allows charge to be transported from one edge to another when flux is inserted - the quantum Hall effect. Hence, spectral flow is preserved over part of the physical spectrum. This conclusion is supported by studies of the level statistics of the physical energy spectrum and entanglement spectrum of the CI.\cite{Prodan2010}  

As $W$ increases, the range of physical energies in which both bulk and edge states display spectral flow is compressed and finally vanishes at $W_c$ - this has been described as `levitation and annihilation'.\cite{Onoda2007} At this point the edge states cease to extend around the cylinder and localize. The trace index vanishes because localized wavefunctions do not disperse with the flux and there are no longer any extended single-particle states that intersect $\varepsilon_{\sma{F}}^{\sma{\Phi}}$ in the mobility gap. 

\begin{figure*}
\centering
\includegraphics[width=16cm]{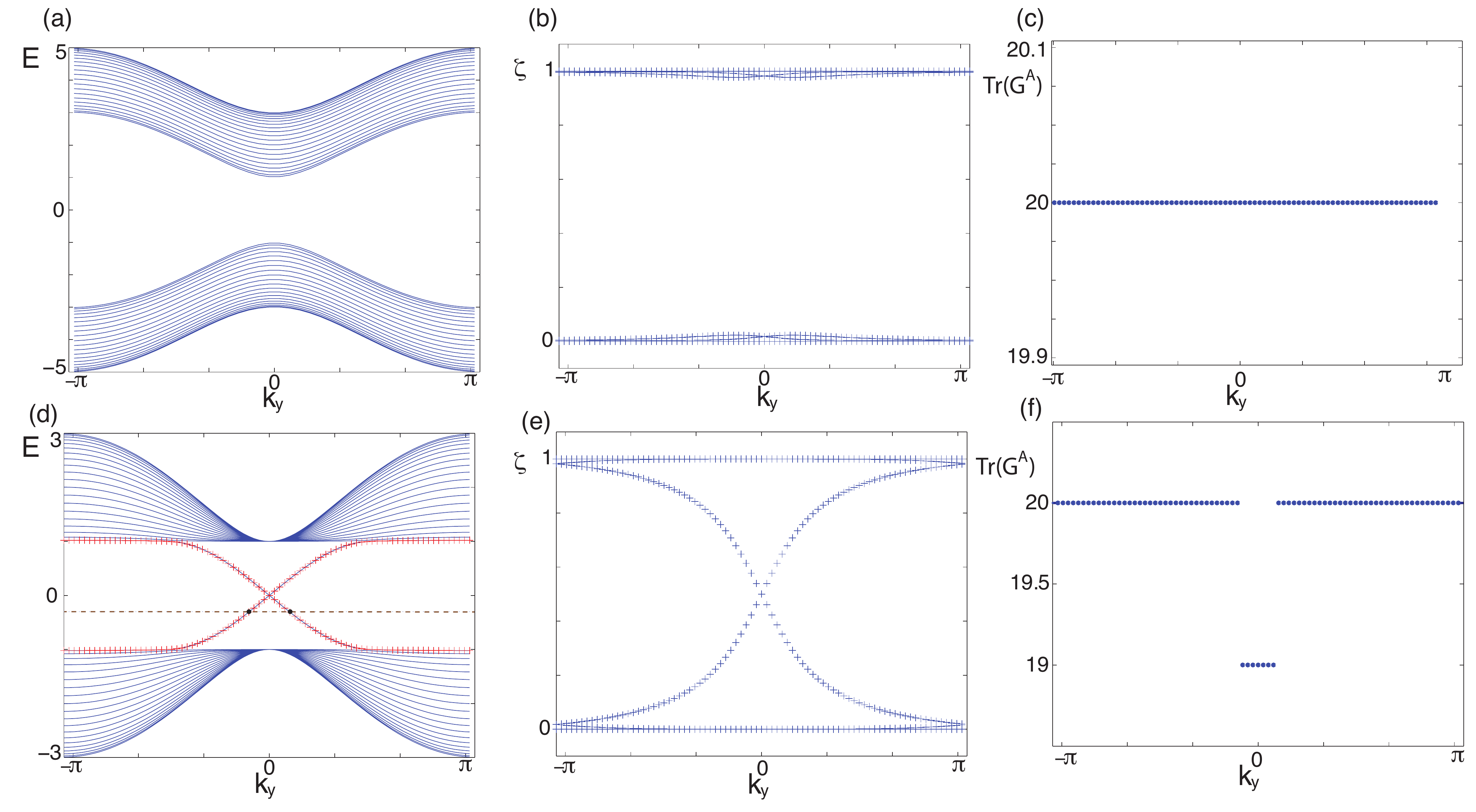}
\caption{Top: Figures \ref{fig:QSHIcompare}-a,b,c  show, respectively, the physical energy spectrum, the single-particle entanglement spectrum (ES), and the trace of $\GA(k_y)$ for a trivial insulator (${\chi}^{\sma{Z_2}}$ even) on an open-end cylinder. In Fig. \ref{fig:QSHIcompare}-a and b, bulk bands on opposite sides of the gap are spectrally disconnected from each other. In Fig. \ref{fig:QSHIcompare}-c, the trace of $\GA$ is a continuous function of $k_y$ so the trace index ${\cal A}_{\sma{Z_2}}=0$. Top: Figures \ref{fig:QSHIcompare}-d,e,f show, respectively, the physical energy spectrum, the single-particle ES, and the trace of $\GA(k_y)$ for a nontrivial insulator with odd ${\chi}^{\sma{Z_2}}$. In Fig. \ref{fig:QSHIcompare}-d and e, a set of edge states interpolate between bulk bands on opposite sides of the gap - this spectral flow distinguishes trivial and nontrivial $Z_2$ insulators. In Fig. \ref{fig:QSHIcompare}-f, the trace of $\GA$ jumps discontinuously by $1$ in half the Brillouin zone $k_y \in [0,\pi]$ - the trace index ${\cal A}_{\sma{Z_2}}=+1$.}\label{fig:QSHIcompare}
\end{figure*}

\section{Quantum Spin Hall Insulator} \label{QSHI}

The quantum spin Hall insulator (QSHI) is a time-reversal invariant topological insulator\cite{kane2005A,kane2005B,bernevig2006a,bernevig2006c,koenig2007} which is most easily modeled by combining two CI's, one for each spin and with opposite chiralities. This insulator is realized in HgTe/CdTe quantum wells and has an effective four-band Hamiltonian on the square lattice with periodic boundary conditions: 
\begin{eqnarray} \label{eq:MofK}
h(k)&=&\sin k_x\hat{\Gamma}_1+\sin k_y\hat{\Gamma}_2+M(k)\hat{\Gamma}_0\label{eq:QSH_H}\\
M(k)&=&2-m-\cos k_x-\cos k_y.
\end{eqnarray}\noindent where $\hat{\Gamma}_1=\sigma_3 \otimes\tau_1, \hat{\Gamma}_2=I_s \otimes\tau_2, \hat{\Gamma}_0=I_s \otimes\tau_3$. Let $I_s$ and $I_o$ be the identities in spin and orbital space respectively; $\sigma^a$ and $\tau^a$ are Pauli matrices in spin and orbital spaces respectively. The time-reversal operator is $T= (i\sigma_2\otimes I_o) K$. It exhibits phases analogous to the CI \emph{i.e.} it is topological for $0<m<4$ when gapped, and trivial otherwise. 
The total Chern number of the ground state vanishes but there is a  $Z_2$ invariant (${\Huge \chi}^{\sma{Z_2}}$) which classifies the two possible equivalence classes of time-reversal symmetric Hamiltonians. We choose the convention that ${\chi}^{\sma{Z_2}}$ is odd (even) when the insulator is nontrivial (trivial).

In our chosen geometry, we numerically plot in Fig. \ref{fig:QSHIcompare}-a and d the physical energy spectra of a ${\chi}^{\sma{Z_2}}$-even and odd insulator respectively. In the ${\chi}^{\sma{Z_2}}$-even trivial case, the conduction and valence bands are spectrally disconnected; in the nontrivial case, the conduction and valence bands are spectrally connected by an interpolating set of edge states. The single-particle entanglement spectra of trivial (Fig. \ref{fig:QSHIcompare}-b) and nontrivial (Fig. \ref{fig:QSHIcompare}-e) insulators are also distinguished by a set of edge states which interpolate across the entanglement gap for the nontrivial case. These edge states extend over the entanglement cut between regions $A$ and $B$ and contribute maximally to the entanglement entropy. 

In principle, one may add any time-reversal symmetric couplings to the Hamiltonian of (\ref{eq:MofK}), including terms that couple the spin-up and spin-down sectors. These terms may change the energy spectrum and the character of the ground-state wavefunction. However, as long as both the symmetry and the energy gap are preserved, the topological invariant ${\chi}^{\sma{Z_2}}$ cannot change. The following discussion of the $Z_2$ trace index is applicable to the most general form of a time-reversal invariant Hamiltonian.

\begin{figure*}
\centering
\includegraphics[width=16.0cm]{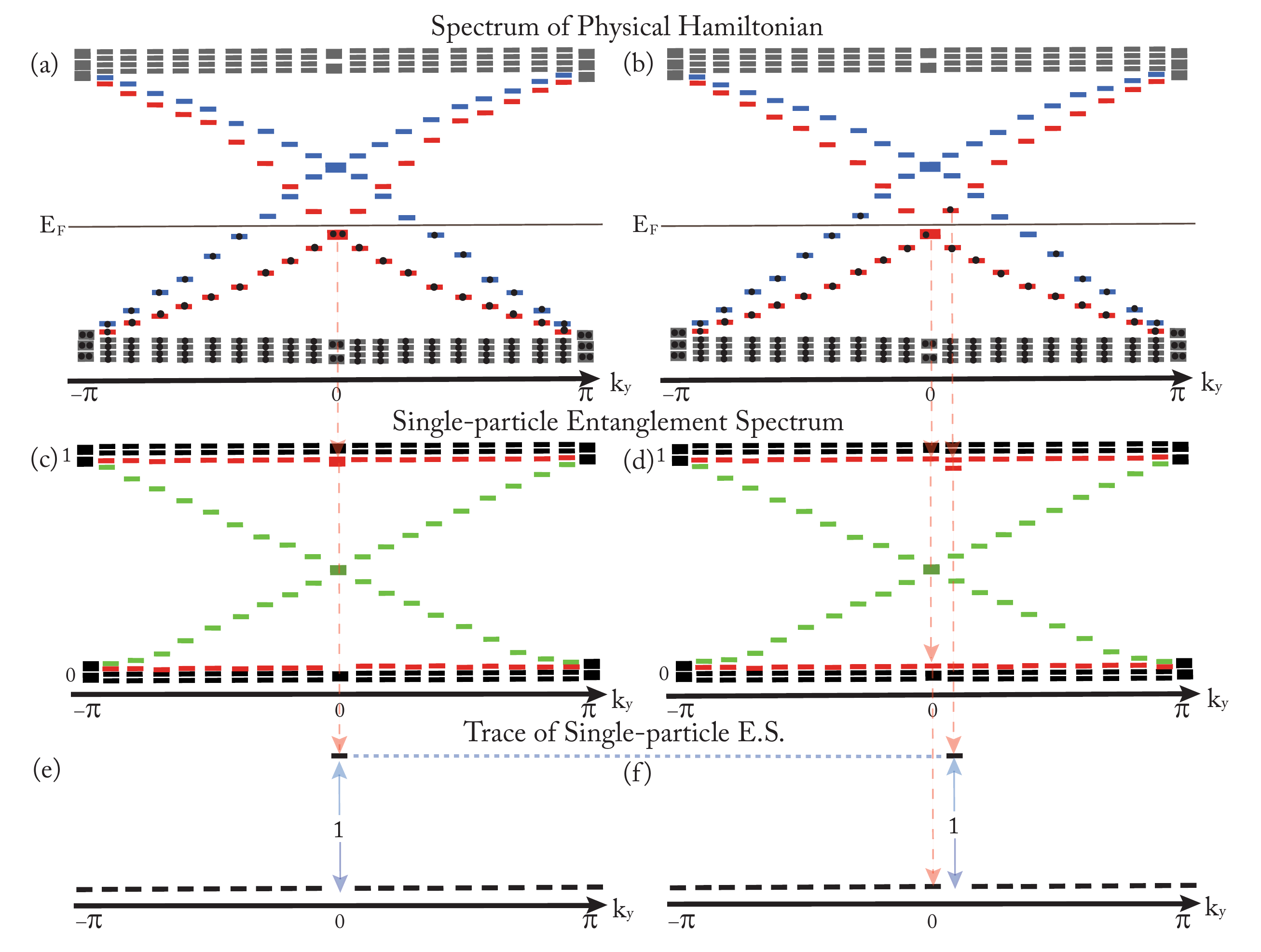}
\caption{Left side: Figures \ref{QSHIschematic1}-a,c,e show schematically the physical energy spectrum, the single-particle entanglement spectrum (ES) and the trace of the single-particle ES respectively, as functions of $k_y$, for a many-body ground state of a nontrivial QSHI. The subspace of occupied bands in the physical spectrum is closed under time-reversal. Right-side: Figures \ref{QSHIschematic1}-b,d,f label the new functions when an integer charge-flux is inserted through the symmetry axis of the cylinder. For convenience, bulk bands away from the gaps are portrayed as dispersionless. Eigenvalues that are colored red, green and blue correspond to eigenstates that extend over the left edge, the entanglement cut and the right edge of the cylinder respectively. Kramers' doublets at $k_y=0$ and $\pi$ are labelled by thick dashes in the physical energy spectra; cccupied single-particle states of a many-body Slater determinant are labelled by black circles. }\label{QSHIschematic1}
\end{figure*}

\begin{figure*}
\centering
\includegraphics[width=16.0cm]{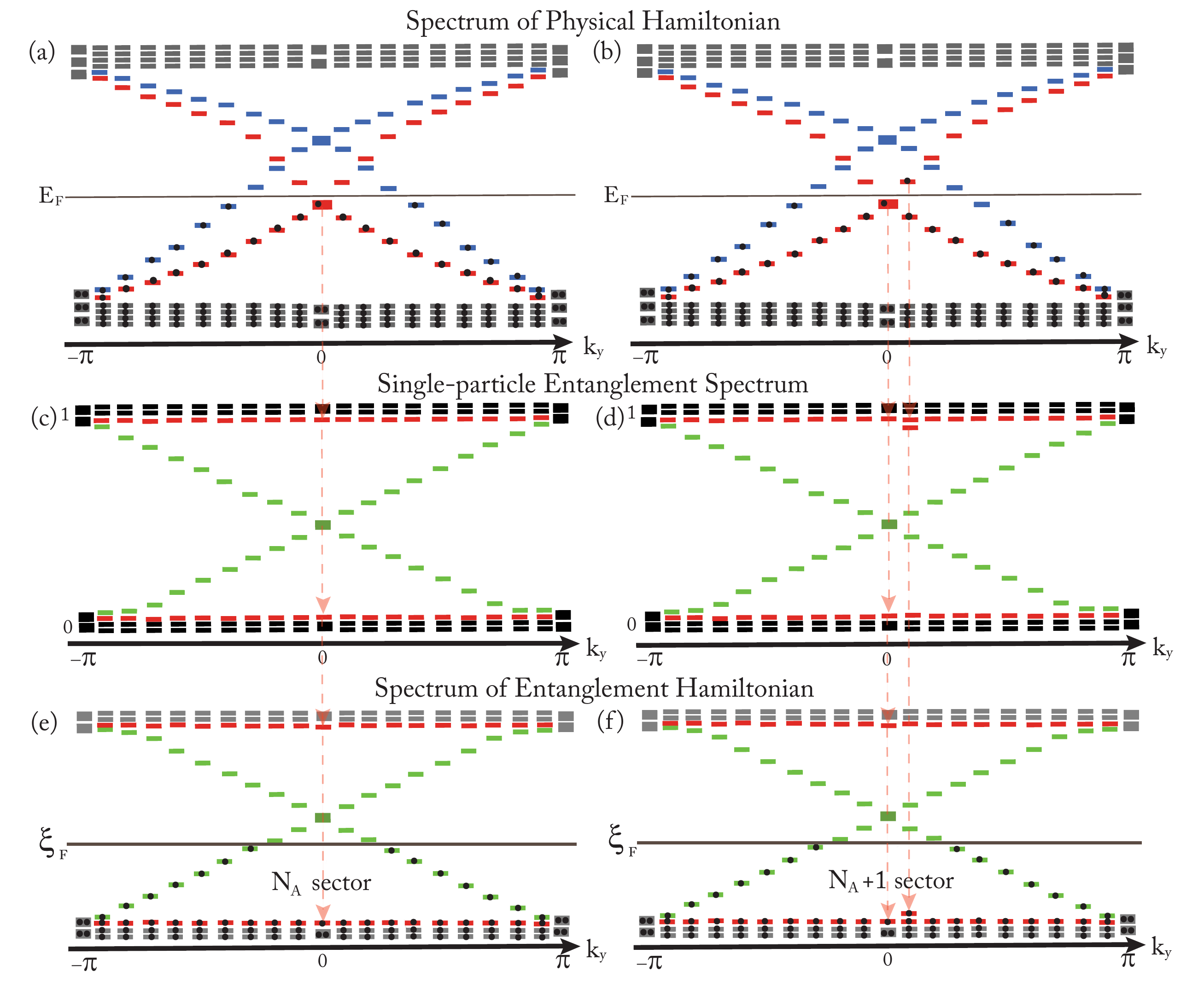}
\caption{Left side: Figures \ref{QSHIschematic2}-a,c,e show schematically the physical energy spectrum, the single-particle entanglement spectrum (ES) and the entanglement energy spectrum respectively, as functions of $k_y$, for a many-body ground state of a nontrivial QSHI. The subspace of occupied bands in the physical spectrum is \emph{not} closed under time-reversal. Right-side: Figures \ref{QSHIschematic2}-b,d,f label the new functions after $\pi$ charge-flux and $\pi$ $\Gamma$-flux are inserted in succession through the symmetry axis of the cylinder. For convenience, bulk bands away from the gaps are portrayed as dispersionless. Eigenvalues that are colored red, green and blue correspond to eigenstates that extend over the left edge, the entanglement cut and the right edge of the cylinder respectively. Kramers' doublets at $k_y=0$ and $\pi$ are labelled by thick dashes in the physical and entanglement energy spectra; cccupied single-particle states of a many-body Slater determinant are labelled by black circles. }\label{QSHIschematic2}
\end{figure*}

\subsection{QSHI: The trace index and its equivalence to the $Z_2$ invariant} \label{z2}

In this section we define the trace index for a $Z_2$ insulator and show its equivalence to the $Z_2$ invariant. The differences from the $U(1)$ trace index are that the $Z_2$ index is defined over half the Brillouin zone and only the parity of the index is physically relevant. 

The correlation matrix $G(k_y)$ as defined in (\ref{cormatrix}) satisfies $T G(k_y) T^{-1} = G(-k_y)$ if the subspace of occupied bands at $\pm k_y$ in the physical spectrum is closed under time-reversal. This condition means that if any single-particle state $\vert\psi\rangle$ is occupied, so is its time-reversed partner  $T\vert\psi\rangle$. We prove this claim in Appendix \ref{cormatrixtrs}. We assume this is true for all $k_y$ and will later generalize our discussion. The reduced correlation matrix $\GA(k_y)$ as defined in (\ref{reducedcormatrix}) is the matrix $G(k_y)$ with spatial indices restricted to region $A$.  Since the time-reversal operation $T = K\, i\sigma_2 \otimes I_o$ is diagonal in spatial indices, a spatial entanglement cut cannot break time-reversal symmetry and $T \GA (k_y) T^{-1} = \GA (-k_y)$ as well. This symmetry allows us to extract all relevant information from half the Brillouin zone ($k_y \in [0,\pi]$). 

The following discussion mirrors that in Section \ref{Chernanomaly}.
Let $K_{\sma{\text{cross}}}^{\sma{(1/2)}}$ be the values of $k_y$ in half the Brillouin zone where states that are localized on  the edge of $A$ intersect the Fermi level. Our first example is a ${\chi}^{\sma{Z_2}}$-odd insulator with one such intersection at $K_{\sma{\text{cross}}}^{\sma{(1/2)}} = \tfrac{\pi}{N_y}$. As illustrated in Fig. \ref{QSHIschematic1}-a, its physical energy spectrum exhibits spectral flow - the conduction and valence bulk bands are spectrally connected across the gap by an interpolating set of edge states. We choose our Fermi energy to lie just above the Kramers' doublet of left-edge states at $k_y=0$, so each occupied single-particle state has a Kramers' partner which is also occupied. Following the discussion in the previous paragraph, we conclude that $T \GA(k_y) T^{-1} = \GA(-k_y)$ for all $k_y$. We plot the time-reversal symmetric spectra of $\GA$ in Fig. \ref{QSHIschematic1}-c.

We define the $Z_2$ trace index (${\cal A}_{\sma{Z_2}}$) for clean, free insulators as the sum of the trace discontinuities over all $K_{\sma{\text{cross}}}^{\sma{(1/2)}}$ in the thermodynamic limit: 
\bal \label{z2ano}
{\cal A}_{\sma{Z_2}} &\equiv\; {\displaystyle \lim_{N_x,N_y \to \infty}} \sum_{K_{\sma{\text{cross}}}^{\sma{(1/2)}}} \text{Tr}\;\big[\GA\big] (K_{\sma{\text{cross}}}^{\sma{(1/2)}} + \tfrac{\pi}{N_y}) \lin
&\;\;\;\;- \text{Tr}\;\big[\GA\big] ( K_{\sma{\text{cross}}}^{\sma{(1/2)}} - \tfrac{\pi}{N_y}) 
\end{align}
As illustrated in Fig. \ref{QSHIschematic1}-e, the trace of $\GA$ decreases discontinuously by $1$ between $k_y=0$ and $\tfrac{2\pi}{N_y}$ in our example, so ${\cal A}_{\sma{Z_2}}=-1$ and has the same parity as ${\chi}^{\sma{Z_2}}$.

More generally, we claim that ${\cal A}_{\sma{Z_2}}$ is modulo $2$ the total number of left-edge states that intersect the Fermi level in half the Brillouin zone. Following the discussion in Section \ref{Chernanomaly} we apply Theorem \ref{thm:1} and obtain ${\cal A}_{\sma{Z_2}} =  n_{\sma{(-)}} - n_{\sma{(+)}}$ where $n_{\sma{(+)}}$ ($n_{\sma{(-)}}$) is the number of left-edge states that intersect the Fermi energy with positive (negative) $\tfrac{\partial \varepsilon}{\partial k_y}$. The claim is proven by the identity $n_{\sma{(-)}} - n_{\sma{(+)}} = n_{\sma{(-)}} + n_{\sma{(+)}}$ mod $2$.  

If $n_{\sma{(-)}} + n_{\sma{(+)}}$ is odd, the physical energy spectrum is characterized by spectral flow, which cannot be removed by any change in the Hamiltonian that simultaneously preserves the gap and time-reversal symmetry - this phase displays topologically protected edge states and has odd $\chi^{\sma{Z_2}}$. The trivial phase corresponds to both $\chi^{\sma{Z_2}}$ and $n_{\sma{(-)}} + n_{\sma{(+)}}$ being even, in which case there is no spectral flow. Since ${\cal A}_{\sma{Z_2}} = n_{\sma{(-)}} + n_{\sma{(+)}}$ mod $2$, we identify the $Z_2$ trace index with the $Z_2$ invariant - ${\cal A}_{\sma{Z_2}} = \chi^{\sma{Z_2}} \;\;\text{mod}\;2$ -  with the caveat that the subspace of occupied bands in the many-body ground state is closed under time-reversal. We numerically plot the trace of $\GA$ for the trivial and nontrivial insulators in Fig. \ref{fig:QSHIcompare}-c and f respectively. The Fermi level of the  nontrivial ground-state is depicted in Fig. \ref{fig:QSHIcompare}-d as a dashed line; the subspace of occupied bands is closed under time-reversal. In the trivial case, Tr $\GA$ is a continuous function of $k_y$ while in the nontrivial case it has a discontinuity of $+1$ in half the Brillouin zone, so we confirm that the parities of ${\cal A}_{\sma{Z_2}}$ and $\chi^{\sma{Z_2}}$ are identical.

We now address the exceptional case of a ground state for which the subspace of occupied bands at the symmetric momenta $k_y^{\sma{sym}}=0,\pi$ is not closed under time-reversal. If at $k_y^{\sma{sym}}$ the Kramers' doublet that extends around the edge of $A$ is singly-occupied, the single-particle entanglement spectrum (ES) is no longer time-reversal symmetric, \emph{i.e.} $T \GA(k_y^{\sma{sym}}) T^{-1} \neq \GA(k_y^{\sma{sym}})$. This is apparent because a singly-occupied doublet contributes one zero eigenvalue and one unit eigenvalue to the single-particle ES at $k_y^{\sma{sym}}$, in accordance with Theorem \ref{thm:1}, and neither of these eigenvalues has a degenerate Kramers' partner. 

As an example, we consider an insulator with an odd number of particles, which implies that the ground state is two-fold degenerate. Our previous example of a many-body ground state in Fig. \ref{QSHIschematic1} has an even number of particles, so we remove a particle $\ket{\psi}$ from the two-fold degenerate subspace at energy $E_{\sma{F}}$ and momentum $k_y=0$. This state $\ket{\psi}$ may be any linear combination of the basis states that span the Kramer's subspace at $k_y=0$. Since $T^2 = -I$, the time-reversed state $T\ket{\psi}$ is always orthogonal to $\ket{\psi}$ and hence lies within the occupied subspace - we have thus shown that the occupied subspace is not closed under the action of time-reversal. As illustrated in Fig. \ref{QSHIschematic2}-a, the resultant Slater determinant has a singly-occupied left-edge Kramers' doublet at $k_y=0$. The occupied left-edge state in the physical spectrum translates to an eigenvalue of $1$ at $k_y=0$ in the single-particle ES of Fig. \ref{QSHIschematic2}-c; its unoccupied Kramer's partner in the physical spectrum translates to an eigenvalue of $0$ in Fig. \ref{QSHIschematic2}-c. There are now two isolated single-particle entanglement states, separated across the entanglement gap, at symmetric momentum $k_y=0$; neither of these two states has a Kramer's partner. Since there is no discontinuous change in the occupation numbers of left-edge states, the trace of $\GA$ is now continuous over the half Brillouin zone and the $Z_2$ trace index is zero. We have thus shown that the parity of  ${\cal A}_{\sma{Z_2}}$ is opposite to that of $\chi^{\sma{Z_2}}$ for an exceptional case. 

We remedy this index by introducing a quantity $\Theta$ which we define as the total number of singly-degenerate entanglement states above the entanglement gap at both $k_y^{\sma{sym}}$. In Appendix \ref{app:definetheta} we remark briefly on how to rigorously define $\Theta$ for even and odd $N_y$. Adding $\Theta$ to ${\cal A}_{\sma{Z_2}}$ gives the correct parity of $\chi^{\sma{Z_2}}$ for any value of the Fermi energy:
\bal
{\cal A}_{\sma{Z_2}} + \Theta = \chi^{\sma{Z_2}} \;\;\text{mod}\;2.
\end{align}
For our exceptional example, Fig. \ref{QSHIschematic2}-c reveals a singly-degenerate entanglement eigenstate with unit eigenvalue above the entanglement gap, so $\Theta=1$ and ${\cal A}_{\sma{Z_2}} + \Theta$ gives the correct parity of $\chi^{\sma{Z_2}}$.

\begin{widetext}

\subsection{QSHI: Flux threading and entanglement}

In this section we investigate the effects of threading flux through the symmetry axis of a QSHI in an open-end cylinder geometry. Unlike the CI, the QSHI is constrained by time-reversal symmetry to have zero Chern number and it does not pump any net charge in response to \emph{charge} flux - flux that induces a spin- and orbital-independent response. For the QSHI, we also consider flux that differentiates electrons with different spin and/or orbitals. The flux threading may be implemented by modifying the hoppings between sites with $y=1$ and $y=N_y$ so that they carry a Peierls's matrix $e^{ i2\pi \Gamma \Phi/\Phi_0} $ with $\Gamma$ a Hermitian matrix in spin-orbital space - we call this \emph{matrix} or $\Gamma$-flux. The tight-binding model in (\ref{eq:2dham}) now has the nearest-neighbor form
\bal \label{eq:matrixfluxHam}
H(\Phi) = \sum_{x y,\alpha\beta\delta} c^{\dagger}_{x y \alpha} h^{\alpha\beta}_{(x,y)(x,y)}c^{\pdag}_{x y\beta} + \bigg(c^{\dagger}_{x+1 y \alpha} h^{\alpha\beta}_{(x+1,y)(x,y)}c^{\pdag}_{x y\beta} + c^{\dagger}_{x y+1 \alpha} h^{\alpha\delta}_{(x,y+1)(x,y)} \big[e^{i 2\pi \Gamma \tfrac{\Phi}{\Phi_0} \delta_{y,N_y}}\big]_{\delta \beta} c^{\pdag}_{x y\beta} + \text{h.c.} \bigg)
\end{align}
with the matrix $h^{\alpha \beta}_{ij}$ obtained from Fourier transforming the QSHI model (\ref{eq:MofK}) in $\hat{x}$ and $\hat{y}$. $\alpha, \beta$ are spin/orbital indices and $i,j,(x,y)$ are lattice coordinates. If the matrices $h_{(1 x)(N_y x)}$ and  $e^{i 2\pi \Gamma \Phi/\Phi_0}$ do not commute, the Hamiltonian (\ref{eq:matrixfluxHam}) cannot be reduced to a flux-free Hamiltonian by a different choice of gauge. In this case, the energy spectrum is changed by inserting matrix flux and translational symmetry along $\hat{y}$ is lost. 
\end{widetext}

  In Section \ref{sec:chargeflux} we investigate the effect of threading charge flux ($\Gamma = I_s \otimes I_o$) on the trace index of the single-particle entanglement spectrum. We show that a certain choice of $\Gamma$ induces charge transport in Section \ref{sec:spinflux} and demonstrate another method to extract the $Z_2$ invariant. In Section \ref{sec:z2manybody} we identify the signature of charge transport in the many-body entanglement spectrum as a flow in particle-number space.   

\subsubsection{Charge flux and time-reversal polarization: effect on entanglement} \label{sec:chargeflux}

Because the QSHI Hamiltonian has global $U(1)$ symmetry and $\Gamma= I_s \otimes I_o$ commutes with $h_{(1 x)(N_y x)}$, the charge-flux-inserted Hamiltonian is gauge-equivalent to a flux-free Hamiltonian. The physical energy spectrum is thus invariant and $k_y$ is still a good quantum number. The effect of a unit charge flux is to transform single-particle states with $k_y$ to adjacent states at $k_y + \tfrac{2\pi}{N_y}$. Unlike the CI, there is no net transfer of charge between edges because every inter-edge current has a time-reversed partner. Nevertheless, the occupation numbers of the single-particle states change due to a quantum spin Hall response.  We illustrate this process for a $\chi^{\sma{Z_2}}$-odd insulator with a many-body ground state depicted in Fig. \ref{QSHIschematic1}-a; after inserting a unit flux, the corresponding flux-evolved state is shown in Fig. \ref{QSHIschematic1}-b. The resultant wavefunction is $\dg{[{\gamma}_{\sma{L}}^{\sma{I}}]} {\gamma}_{\sma{L}}^{\sma{II}} \dg{[{\gamma}_{\sma{R}}^{\sma{I}}]} {\gamma}_{\sma{R}}^{\sma{II}}\ket{\psi_{\sma{GS}}}$ where ${\gamma}_{\sma{R}}^{\sma{II}}$ and ${\gamma}_{\sma{R}}^{\sma{I}}$ are normal mode operators of the lowest energy states on the right edge and $I,II$ denote time-reversed partner modes. 

As with the Chern insulator, the positions of the discontinuities in the trace are translated by $K_{\sma{\text{cross}}}^{\sma{(1/2)}} \rightarrow K_{\sma{\text{cross}}}^{\sma{(1/2)}} + \tfrac{2\pi}{N_y}$ but the magnitudes of the discontinuities are invariant. Since the $Z_2$ trace index ${\cal A}_{\sma{Z_2}}$ is defined over half the Brillouin zone ($k_y \in [0,\pi]$), it is possible to (i) translate a discontinuity at $K_{\sma{\text{cross}}}^{\sma{(1/2)}} = \pi- \tfrac{\pi}{N_y}$ out of the half-BZ or (ii) translate a discontinuity at $-\tfrac{\pi}{N_y}$ into the half-BZ. For our example,  the discontinuity of $+1$ at $-\tfrac{\pi}{N_y}$ in Fig. \ref{QSHIschematic1}-e (before) is translated into the half-BZ at  $+\tfrac{\pi}{N_y}$ in Fig. \ref{QSHIschematic1}-f (after). Before, ${\cal A}_{\sma{Z_2}}=-1$; after, the discontinuities at $\tfrac{\pi}{N_y}$ and $\tfrac{3\pi}{N_y}$ cancel so  ${\cal A}_{\sma{Z_2}}=0$.

We have shown that the index ${\cal A}_{\sma{Z_2}}$ is not invariant under flux insertion. However, (i) is always accompanied by a change in parity of the occupancy of the $k_y=\pi$ left-edge Kramers' doublet (in the physical spectrum) and similarly (ii) changes the parity at $k_y=0$. The resultant change in parity of $\Theta$ in the entanglement spectrum compensates the change in parity of ${\cal A}_{\sma{Z_2}}$ so that the net parity is invariant.  For our example, we may compare the spectra of $\GA$ in Fig. \ref{QSHIschematic1}-c (before flux) and d (after). The difference is that the flux-evolved spectra of $\GA$ has an additional  non-degenerate eigenvalue of $1$ at $k_y=0$, thus flux-insertion has changed the parity of $\Theta$. 

In Section \ref{z2} we identified ${\cal A}_{\sma{Z_2}} + \Theta$ with the $Z_2$ invariant of the ground state; in this Section we show that this equality may be extended to any flux-inserted state. This implies that we may separately extract $\chi^{\sma{Z_2}}$ from both the ground-state and its flux-inserted state through ${\cal A}_{\sma{Z_2}} + \Theta$.

\subsubsection{Matrix flux and charge transport: effect on entanglement} \label{sec:spinflux}

In a generalization of the Hall effect for $Z_2$ insulators, Qi and Zhang proposed that if $\Gamma$ is odd under time-reversal ($T\,\Gamma\,T^{-1} = -\Gamma$) and satisfies $e^{i\pi \Gamma} = -I$, inserting the matrix flux $e^{i2\pi \Phi/\Phi_0 \Gamma}$ induces charge transport between edges.\cite{qi2008A} Henceforth, we call flux that satisfies these conditions as $\Gamma$-flux. The first condition ensures that the Peierls's matrix satisfies $T \, e^{ i2\pi \Gamma \Phi/\Phi_0}  \,T^{-1} = e^{ i2\pi \Gamma \Phi/\Phi_0} $ so the Peierls-modified Hamiltonian is time-reversal symmetric for all values of $\Phi$. The second condition implies that for special values of the flux ($\Phi = n \Phi_0/2$ with $n$ an integer), the matrix $e^{i 2\pi \Gamma \Phi/\Phi_0}$ is proportional to the identity and commutes with $h_{(1 x)(N_y x)}$; $\pi$ ($2\pi$) $\Gamma$-flux has the same effect as $\pi$ $(2\pi)$ charge flux on the physical spectrum.

A choice of $\Gamma$ that satisfies both conditions is $\hat{n}_i \sigma_i \otimes I_o$ where $\hat{n}$ is a unit vector. This choice is called spin flux because it picks a preferred direction ($\hat{n}$) in spin space. Another valid choice, $\Gamma = I_s \otimes \tau_2$, picks a preferred direction in orbital space. The authors in Ref. \onlinecite{qi2008A} showed that different choices of $\Gamma$ cannot change the parity of the charge pumped during an adiabatic insertion of $\pi$ of $\Gamma$-flux; this parity is precisely the $Z_2$ invariant $\chi^{\sma{Z_2}}$.

\begin{figure}
\centering
\includegraphics[width=8.5cm]{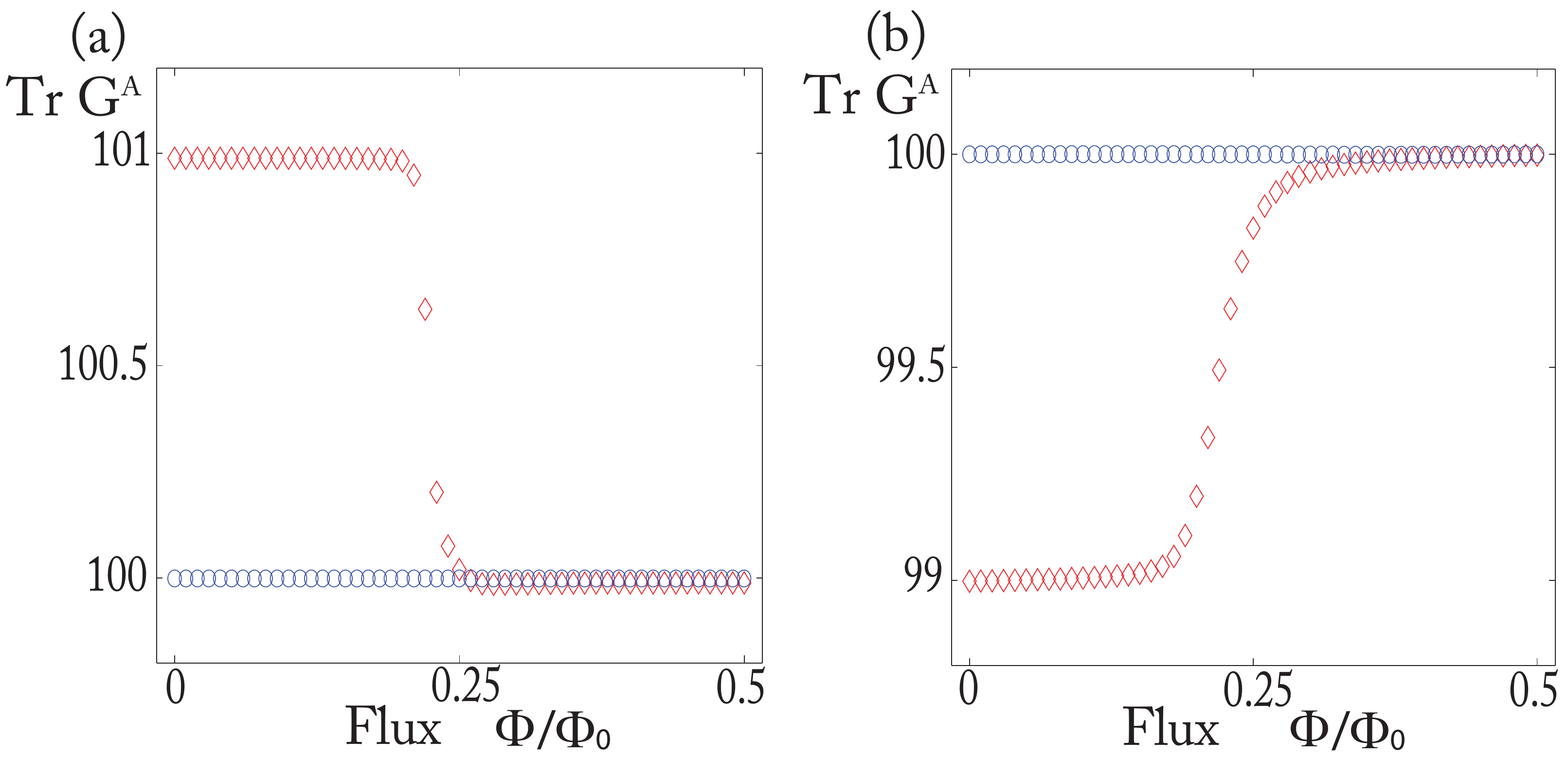}
\caption{ Trace of $\GA$ vs flux $\Phi$ for a trivial $\chi^{\sma{Z_2}}$-even insulator (red) and a nontrivial $\chi^{\sma{Z_2}}$-odd insulator (blue). Flux is threaded by inserting a  matrix $e^{i2\pi \Gamma \Phi/\Phi_o}$ at nearest-neighbor hoppings between $y=10$ to $y=1$ on a $10 \times 10$ lattice. Our Hamiltonian is Eq. (\ref{eq:MofK}) with (i) an additional inversion-symmetry-breaking term $0.1 \si k_x \sigma_1 \otimes I_o + 0.1 \si k_y \sigma_2 \otimes \tau_3$ that couples spin-up and spin-down sectors  and (ii) a weak random on-site potential so that the insulator is disordered; $\chi^{\sma{Z_2}}$ is odd (even) when $m = 1$ ($-1$). In Fig. \ref{fig:tracevsmatrixflux}-a (b) we choose $\Gamma = \sigma_1 \otimes I_o$ ($I_s \otimes \tau_2$). For both choices of $\Gamma$, the trace of $\GA$ smoothly changes by $1$ as $\pi$ $\Gamma$-flux is threaded through the nontrivial insulator. In the trivial case, the traces at $\pi$ and $0$ flux are equal.  }\label{fig:tracevsmatrixflux}
\end{figure}

We now present a way to extract $\chi^{\sma{Z_2}}$ from two many-body states differing by $\pi$ $\Gamma$-flux through the entanglement spectrum; this method applies to disordered and/or interacting $Z_2$ insulators as well. In Section \ref{section:fluxChern} we showed that the full trace of the reduced correlation matrix is the number of particles in region $A$, \emph{i.e.} $ \text{Tr} \;G^{\sma{A}} =  \bra{\Psi}{\hat{N}}_{A} \ket{\Psi}$. Let $\GA_{\sma{0}}$ and $\GA_{\sma{\pi \Gamma}}$ respectively denote the reduced correlation matrices before and after insertion of  $\pi \, \Gamma$-flux.
The difference in the trace of these two matrices is the change in the number of particles in region $A$, which is also the charged pumped when $\pi$ $\Gamma$-flux is threaded:
 \bal \label{eq:pump}
\text{Tr}\;\GA_{\sma{\pi \Gamma}} - \text{Tr} \;\GA_{\sma{0}} = \chi^{\sma{Z_2}} \;\;\text{mod}\;2.
\end{align}
In Fig. \ref{fig:tracevsmatrixflux}-a we thread spin flux ($\Gamma = \sigma_1 \otimes I_o$) separately through a trivial and nontrivial disordered QSHI and plot the trace of $\GA$ as a function of flux $\Phi$ for each case. The result is that $\text{Tr}\;\GA_{\sma{\pi \Gamma}} - \text{Tr} \;\GA_{\sma{0}}=1$ for the nontrivial insulator with odd $\chi^{\sma{Z_2}}$ and equals  $0$ for the trivial insulator with even $\chi^{\sma{Z_2}}$. If instead we thread orbital flux ($\Gamma = I_s \otimes \tau_2$), we show in Fig. \ref{fig:tracevsmatrixflux}-b that no charge is pumped for the trivial insulator and $\text{Tr}\;\GA_{\sma{\pi \Gamma}} - \text{Tr} \;\GA_{\sma{0}}=-1$ for the nontrivial insulator. In either case, the parities of the pumped charges do not change with a different choice of $\Gamma$.

\subsubsection{Spectral flow and the many-body entanglement spectrum} \label{sec:z2manybody}

In this section we investigate how spectral flow manifests itself in the many-body entanglement spectrum (ES) of a $Z_2$ insulator with a cylinder geometry. Since $e^{i \pi \Gamma}=-I$, inserting $\pi$ $\Gamma$-flux has the same effect on the physical spectrum as inserting $\pi$ charge flux. To make a closed loop in parameter space we thread in succession $\pi$ $\Gamma$-flux and $\pi$ charge flux ($\pi \Gamma \pi I$). Since the $Z_2$ insulator is time-reversal symmetric, charge is pumped only in the first half of the loop when we thread $\Gamma$-flux. The parity of the \emph{net} charge that is pumped in this closed loop is thus the $Z_2$ invariant. Let us define the amount of charge pumped from the right to left edge in this adiabatic loop by $Q_{\Gamma}$. The absolute value of $Q_{\Gamma}$ depends on the choice of $\Gamma$, however we have shown that $Q_{\Gamma} = \chi^{\sma{Z_2}}$ mod $2$. 

The following discussion is analogous to the discussion of the CI in Sec. \ref{sect:specflowmanybody}.
Let $\ket{\Psi_{\sma{0}}}$ and $\ket{\Psi_{\sma{\pi \Gamma \pi I}}}$ be the many-body wavefunctions before and after the insertion respectively. The physical energy spectrum is invariant under $\pi \Gamma \pi I$ flux but the number of occupied edge states in region $A$ increases by $Q_{\Gamma}$. As an example we thread $\pi \Gamma \pi I$ flux into the many-body ground state of a $\chi^{\sma{Z_2}}$-odd insulator as depicted in Fig. \ref{QSHIschematic2}-a. We may pick $\Gamma$ such that $Q_{\Gamma}=1$; the resultant wavefunction is shown in Fig. \ref{QSHIschematic2}-b to have an additional left-edge particle at $k_y=\tfrac{2\pi}{N_y}$. We compare the entanglement energy spectra before (Fig. \ref{QSHIschematic2}-e) and after (\ref{QSHIschematic2}-f) flux threading; the difference is that the flux-excited insulator has an additional eigenvalue $\xi = -\infty$ at $k_y=\tfrac{2\pi}{N_y}$. This means that if we began with an entanglement ground state with $N_{\sma{A}}$ particles under a Fermi level $\xi_{\sma{F}}$, the flux-evolved entanglement ground state now has $N_{\sma{A}}+1$ particles under the same Fermi level $\xi_{\sma{F}}$. In Sec. \ref{sect:specflowmanybody} we argued that the extra entanglement energy at $\xi = -\infty$ cannot change the entanglement entropy, so the $N_{\sma{A}}$ sector of the many-body ES before threading is identical to the $N_{\sma{A}}+1$ sector after threading.

More generally, we apply Theorem \ref{thm:1} and the proof in Appendix \ref{app:n0n1} to derive a flow between particle number sectors:
\bal \label{eq:Z2flow}
\bigg\{\lambda \bigg\}_{N_{\sma{A}}}^{\ket{\Psi_{0}}} \eq \bigg\{\lambda \bigg\}_{N_{\sma{A}}+ Q_{\Gamma}}^{\ket{\Psi_{\sma{\pi \Gamma \pi I}}}}. 
\end{align}
This is the signature of spectral flow in the many-body ES of a $Z_2$ insulator. For interacting insulators that preserve the global $U(1)$ symmetry, this relation is valid as well. We note that the parity of the number sector is changed if the QSHI is nontrivial and unchanged if trivial.
  
Let us define the even (odd) particle-number sector as the set of non-zero eigenvalues of $\rho_{\sma{A}}$ obtained by occupying the single-particle entanglement eigenstates $\zeta_i$ of Eq. (\ref{eq:manybody}) in all possible combinations such that the total occupation number is even (odd):
\bal
\bigg\{\lambda \bigg\}_{\text{even (odd)}} \eq \bigg\{ \;\bigg\{\lambda \bigg\}_{N_{\sma{A}}}\;\bigg| \;N_{\sma{A}} \;\text{even (odd)} \bigg\}. 
\end{align}
As a corollary to (\ref{eq:Z2flow}), we deduce that inserting $\pi \Gamma \pi I$ flux through a nontrivial $Z_2$ insulator results in the even number sector flowing into the odd number sector (and vice versa), so
\bal
\bigg\{\lambda \bigg\}_{\text{even (odd)}}^{\ket{\Psi_{0}}} \;\stackrel{\chi^{\sma{Z_2}} \text{odd}}{=} \;\;\bigg\{\lambda \bigg\}_{\text{odd (even)}}^{\ket{\Psi_{\sma{\pi \Gamma \pi I}}}}.
\end{align}
This is distinguished from the trivial $Z_2$ insulator, for which the parity of the number sector is invariant under the same flux insertion. 


\section{Conclusion} \label{conclusion}

The fundamental property that distinguishes free-fermion topological insulators which produce a quantum Hall response is spectral flow, which is manifest in both the physical energy spectrum and the single-particle entanglement spectrum (ES) for a spatial entanglement cut. By a quantum Hall response, we refer to both time-reversal breaking Chern insulators and time-reversal invariant $Z_2$ insulators. For an open-end Laughlin cylinder geometry, spectral flow is identified by the presence of a set of edge states that interpolate across the energy gap in both physical and entanglement spectra, such that the conduction and valence bands are spectrally connected.

In principle, there are many ways to partition the Hilbert space in an entanglement cut; we would like a cut that explicitly demonstrates the connection between entanglement and quantum Hall physics.
In this article we have found that a spatial entanglement cut that separates the two edges of the cylinder is ideally suited to probing the quantum Hall response of a topological insulator.  During the adiabatic insertion of an appropriate flux, the quantum Hall current  flows directly from one edge of the cylinder to the opposite edge, or equivalently from one subsystem of the entanglement cut to the other. Since the quantities of pumped charge for both Chern and $Z_2$ insulators are directly related to their respective topological invariants, it must be that these topological invariants are encoded in the ES.

In this work we have shown that the signatures of spectral flow in the ES lie in  (i) a non-vanishing `trace index' of the single-particle ES (ii) the adiabatic evolution of the full trace of the single-particle ES when flux is threaded  (iii) a global translation in the particle-number sectors of the many-body ES when flux is threaded. Any of the three signatures provides a means to extract the desired topological invariants. While all three are generalizable to disordered insulators, only (ii) and (iii) are applicable to interacting insulators. \\

\emph{Acknowledgements}
We thank L. Fidkowski for a useful discussion. AA was supported by the First Year Fellowship at Princeton University and NSF CAREER DMR-095242. TLH acknowledges support from the NSF, under grants DMR 0758462 at the University of Illinois, and by the ICMT. BAB was supported by Princeton Startup Funds, NSF CAREER DMR-095242, ONR - N00014-11-1-0635, Darpa - N66001-11-1-4110 and  NSF-MRSEC DMR-0819860 at Princeton University. BAB and TLH also thank Microsoft Station Q for generous hosting during the beginning of this work. BAB thanks Technion, Israel, for generous hosting during the latter stages of this work.

\begin{widetext}
\appendix

\section{Derivation of Eq. \ref{eigrel}}\label{app:Grho}

The one-body reduced correlation matrix as defined in (\ref{reducedcormatrix}) is
\bal \label{yala}
G_{ij}^{\sma{A}}(k_y) \eq \text{Tr}_{\sma{A},k_y}\bigg[ \rho_{\sma{A}}^{\sma{\text{ins}}}(k_y) \dg{c_{k_yi}} c_{k_yj} \bigg] = \sum_{ql} u_i^{q\ast}(k_y)\;u_j^{l}(k_y)\; \text{Tr}_{\sma{A},k_y}\bigg[ \prod_{i=1}^{Q_{\sma{A}}} \frac{e^{-\xi_i(k_y) \dg{\chi_{k_yi}} \chi_{k_yi}}}{1+ e^{-\xi_i(k_y)}}  \dg{\chi_{k_yq}} \chi_{k_yl} \bigg]
\end{align}
Let $\ket{\phi_{k_y}} = \ket{n_{1 k_y} \ldots n_{Q_{\sma{A}}k_y}}$ label a Slater determinant in the momentum sector $k_y$ with the corresponding occupation numbers $\{n_{i k_y}\}$. We need the identity
\bal \label{proof}
&\text{Tr}_{\sma{A},k_y}\bigg[ \prod_{i=1}^{Q_{\sma{A}}} e^{-\xi_i(k_y) \dg{\chi_{k_yi}} \chi_{k_yi}} \dg{\chi_{k_yq}} \chi_{k_yl} \bigg] = \sum_{\phi_{k_y}(1) \phi_{k_y}(2)} \bra{\phi_{k_y}(1)} \prod_{i=1}^{Q_{\sma{A}}} e^{-\xi_i(k_y) \dg{\chi_{k_yi}} \chi_{k_yi}} \ket{\phi_{k_y}(2)}\bra{\phi_{k_y}(2)} \dg{\chi_{k_yq}} \chi_{k_yl} \ket{\phi_{k_y}(1)} \lin
\eq \prod_{i=1}^{Q_{\sma{A}}} \sum_{\phi_{k_y}(1) \phi_{k_y}(2)}^1 e^{- \xi_i(k_y) n_{i k_y}(2)} \bra{ \phi_{k_y}(1)} \phi_{k_y}(2) \rangle  \bra{ \phi_{k_y}(2)} \dg{\chi_{k_yq}} \chi_{k_yl} \ket{\phi_{k_y}(1)} \lin
\eq \prod_{i=1}^{Q_{\sma{A}}} \sum_{n_{i k_y}=0}^1 e^{- \xi_i(k_y) n_{i k_y}} \bra{n_{1 k_y} \ldots n_{\sma{Q_A} k_y}} \dg{\chi_{k_yq}} \chi_{k_yl} \ket{n_{1 k_y} \ldots n_{\sma{Q_A} k_y}} = \prod_{i=1}^{Q_{\sma{A}}}  \sum_{n_{i k_y }=0}^1 e^{- \xi_i(k_y) n_i(k_y)}   \delta_{q,l} \delta_{n_{q k_y},1} \lin 
\eq \delta_{q,l} \;e^{-\xi_q(k_y)} \prod_{i=1,i\neq q}^{Q_{\sma{A}}} \bigg(1+ e^{-\xi_i(k_y)}\bigg).
\end{align}
The second equality follows because $ e^{-\xi_i(k_y) \dg{\chi_{k_yi}} \chi_{k_yi}}$ is a counting operator that is diagonal in the basis $\ket{\phi_{k_y}}$.\end{widetext}
Substitution of this identity in (\ref{yala}) reveals 
\bal
G_{ij}^A(k_y) \eq \sum_{q} v_i^{q\ast}(k_y)\;v_j^{q}(k_y)\; \frac{1}{1+ e^{\xi_q(k_y)}}
\end{align}
Recall that the entanglement Hamiltonian matrix is
\bal
\big[H_{\sma{\text{ent}}}\big]_{ij}(k_y) \eq \sum_q v_i^{q}(k_y)\;v_j^{q\ast}(k_y)\; \xi_q(k_y)
\end{align}
Hence Eq. (\ref{eigrel}).

\section{Proof of Theorem \ref{thm:1}} \label{app:thm1}
\begin{proof}
From Eq. (\ref{cormatrix}), we know that the complex conjugate of the correlation matrix $G$ is a sum of projectors of single-particle eigenstates of the Hamiltonian $H$ which are occupied in the many-body wavefunction. Let us define $G$ as the correlation matrix of a many-body wavefunction that includes the single-particle state $u^{(L)}$. Since $u^{(L)}$ is an eigenstate of $H$ we may form a projector matrix $P^{(L)}_{ij} = u^{(L)}_iu^{(L)\ast}_j$ and define $\bar{G}$ as the difference between $G$ and $P^{(L)\ast}$
\bal
\bar{G} \eq G - P^{(L)\ast}
\end{align}
$\bar{G}$ is the correlation matrix of a many-body wavefunction that excludes the single-particle state $u^{(L)}$.
We define $\bar{G}^{\sma{A}}$ and $P^{(L)A}$ respectively as the matrices $\bar{G}$ and $P^{(L)}$ with indices restricted to $A$. 
Since $\bar{G}^{\ast}$ is a sum of projectors of states orthogonal to $u^{(L)}$, 
\bal
\forall i \in A \otimes B,\;\;\sum_{j \in A \otimes B}\bar{G}_{ij} u^{(L)\ast}_j \eq 0
\end{align}
Since $u^{(L)}$ is confined in $A$, $u^{(L)}_j =0$ if $j \in B$. This means that 
\bal \label{eq:zeroeig}
\forall i \in A \otimes B,\;\;\sum_{j \in A}\bar{G}_{ij} u^{(L)\ast}_j \eq 0 \lin
\forall i \in A,\;\; \sum_{j \in A}\bar{G}_{ij} u^{(L)\ast}_j \eq 0 \lin
\forall i \in A,\;\; \sum_{j \in A}\bar{G}^{\sma{A}}_{ij} u^{(L)\ast}_j \eq 0
\end{align}
Hence we have shown that $u^{(L)\ast}$ is an eigenvector of $\bar{G}^{\sma{A}}$ with zero eigenvalue. 
By definition,
\bal \label{eq:proof}
G^{\sma{A}}_{ij} \eq \bar{G}^{\sma{A}}_{ij} + P^{(L)A\ast}_{ij} = \bar{G}^{\sma{A}}_{ij} + u^{(L)\ast}_{i}u^{(L)}_{j}
\end{align}
$P^{(L)A \ast}$ is the projector of the eigenstate $u^{(L)\ast}$ which we have shown to be a zero-eigenvalue eigenvector of $\bar{G}^{\sma{A}}$. In other words, $u^{(L)\ast}$ is orthogonal to all the eigenvectors of $\bar{G}^{\sma{A}}$. Since $G$ is Hermitian, so is $G^{\sma{A}}$; this implies that $G^{\sma{A}}$ is orthogonally diagonalizable. If $G^{\sma{A}}$ is an $n \times n$ matrix, we can find $n$ linearly independent and orthonormal eigenvectors $v$ that diagonalizes $G^{\sma{A}}$. Let $\lambda^{(i)}$ denote the eigenvalues of  $G^{\sma{A}}$ with corresponding eigenvector $v^{(i)}$, then 
\bal
G^{\sma{A}}_{ij} \eq \sum_{q=2}^n \lambda^{(q)} v^{(q)}_{i} v^{(q)\ast}_{j} + \lambda^{(1)} v^{(1)}_{i} v^{(1)\ast}_{j}
\end{align}
with $v^{(1)}$ orthogonal to the matrix $\sum_{i=2}^n \lambda^{(i)} v^{(i)}_{i} v^{(i)\ast}_{j} $. In fact, we have already found such a form for $G^{\sma{A}}$ in Eq. (\ref{eq:proof}) and we need only identify the eigenvector $v^{(1)}$ as $u^{(L)\ast}$ and the eigenvalue $\lambda^{(1)}$ as unity. \qed
\end{proof}

\section{Proof of invariance of the $U(1)$ trace index under integer flux insertion} \label{app:u1inv}

For a generic CI with Chern number $C_1$, the left-edge states intersect the Fermi energy $n_{+}$ times with $\tfrac{\partial \varepsilon}{\partial k_y}>0$ and $n_{-}$ times with $\tfrac{\partial \varepsilon}{\partial k_y}<0$, such that $n_{-}-n_{+} = C_1$. There are $n_{+}$ values of $K_{\sma{\text{cross}}}^+$ where the trace of $G^{\sma{A}}$ discontinuously decreases by $1$ and $n_{-}$ values of $K_{\sma{\text{cross}}}^-$ where the trace of $G^{\sma{A}}$ discontinuously increases by $1$.

 According to Theorem \ref{thm:1}, for each $K_{\sma{\text{cross}}}^+$ there is at least one occupied left-edge state at $K_{\sma{\text{cross}}}^+ - \tfrac{\pi}{N_y}$ which is a unit-eigenvalue eigenstate of $G^{\sma{A}}(K_{\sma{\text{cross}}}^+ - \tfrac{\pi}{N_y})$ and at least one unoccupied left-edge state at $K_{\sma{\text{cross}}}^+ + \tfrac{\pi}{N_y}$ which is a zero-eigenvalue eigenstate of $G^{\sma{A}}(K_{\sma{\text{cross}}}^+ + \tfrac{\pi}{N_y})$ - each of these crossings contributes $-1$ to the trace index. For all $K_{\sma{\text{cross}}}^-$ there is at least one unoccupied left-edge state at $K_{\sma{\text{cross}}}^- - \tfrac{\pi}{N_y}$ which is a zero-eigenvalue eigenstate of $G^{\sma{A}}(K_{\sma{\text{cross}}}^- - \tfrac{\pi}{N_y})$ and at least one occupied left-edge state at $K_{\sma{\text{cross}}}^- + \tfrac{\pi}{N_y}$ which is a unit-eigenvalue eigenstate of $G^{\sma{A}}(K_{\sma{\text{cross}}}^- + \tfrac{\pi}{N_y})$ - each such crossing contributes $+1$ to the trace index. 
 
 Upon threading a unit flux, all states transform from momentum $k_y$ to $k_y + \tfrac{2\pi}{N_y}$. We assume that the flux is small enough that it does not change the occupation numbers of bulk extended states, to keep the phase insulating. For each $K_{\sma{\text{cross}}}^+$, the left-edge state at $K_{\sma{\text{cross}}}^+ + \tfrac{\pi}{N_y}$ is now occupied and is a unit-eigenvalue eigenstate of $G^{\sma{A}}(K_{\sma{\text{cross}}}^+ + \tfrac{\pi}{N_y})$. In the same left-edge mode, the left-edge state at $K_{\sma{\text{cross}}}^+ + \tfrac{3\pi}{N_y}$ is still unoccupied and is a zero-eigenvalue eigenstate of $G^{\sma{A}}(K_{\sma{\text{cross}}}^+ + \tfrac{3\pi}{N_y})$. For each $K_{\sma{\text{cross}}}^+$, the trace of $G^{\sma{A}}$ is now continuous between $k_y=K_{\sma{\text{cross}}}^+ - \tfrac{\pi}{N_y}$ and $K_{\sma{\text{cross}}}^+ + \tfrac{\pi}{N_y}$; there is now a discontinuity of $-1$ in the trace of $G^{\sma{A}}$ between $k_y=K_{\sma{\text{cross}}}^+ + \tfrac{\pi}{N_y}$ and $K_{\sma{\text{cross}}}^+ + \tfrac{3\pi}{N_y}$. 
 
 Analogously for the negative-$\tfrac{\partial \varepsilon}{ \partial k_y}$ crossings, there is now a discontinuity of $+1$ in the trace of  $G^{\sma{A}}$ between $k_y=K_{\sma{\text{cross}}}^- + \tfrac{\pi}{N_y}$ and $K_{\sma{\text{cross}}}^- + \tfrac{3\pi}{N_y}$. 
 
 We have shown that the net result of integer-flux insertion is that all trace discontinuities are translated by $\tfrac{2\pi}{N_y}$. The $U(1)$ trace index is the sum of all these discontinuities and is therefore invariant under integer flux insertion.

\section{Proof of Eq. (\ref{specflowentspec}) and (\ref{eq:Z2flow})}\label{app:n0n1}

According to Theorem \ref{thm:1}, occupied left-edge states are eigenstates of the reduced correlation matrix $G^{\sma{A}}$ with eigenvalue $1$ with finite-size exponential corrections while unoccupied left-edge states have eigenvalue $0$. We define $n_0$ ($n_1$) as the number of eigenvalues that flow exponentially to $0$($1$) in the thermodynamic limit, and $n_{1/2}$ as the number of every other eigenvalue, which we label by $\zeta^{\sma{(1/2)}}_i$. We define $n^{\sma{(1/2)}}_i$ as the occupation numbers of the single-particle particle eigenstates with eigenvalues $\zeta^{\sma{(1/2)}}_i$. By definition, $d = n_0 + n_1 + n_{1/2}$.

All eigenvalues $\lambda$ in the many-body entanglement spectrum are proportional to the factor $\prod_{i \in \text{unoccupied}} (1 - \zeta_i)$ - any nonzero eigenvalue $\lambda$ must arise from occupying all $n_1$ single-particle states with eigenvalue $1$, so the minimum particle-number sector is $n_1$.  

$\lambda$ is also proportional to the factor  $\prod_{i \in \text{occupied}} \zeta_i$. This means we cannot have a particle-number sector larger than $n_{1/2} + n_1$ because that would involve occupying at least one zero-eigenvalue single-particle state.

For a number sector $n_{1} + m$ which falls in the range bounded by $n_1$ and $n_1 + n_{1/2}$, we have from Eq. (\ref{balls}) that
\bal \label{suzy}
\bigg\{ \lambda \bigg\}_{n_1+m} \eq \bigg\{\prod_{i=1}^{n_1+m} \zeta_i \prod_{j=1}^{d-n_1-m} \big(1-\zeta_j \big) \;\bigg| \sum_{i} n_i =n_1+m \bigg\}
\end{align}
with the $i$ ($j$) labelling occupied (unoccupied) entanglement states.
For $\lambda$ to be nonzero, all $n_1$ unit-eigenvalue single-particle states are occupied and all $n_0$ zero-eigenvalue states are unoccupied. This results in
\bal \label{impi}
\prod_{i=1}^{n_1+m} \zeta_i \eq \prod_{j=1}^{n_1} 1 \prod_{k=1}^{m} \zeta_k^{\sma{(1/2)}} = \prod_{k=1}^{m} \zeta_k^{\sma{(1/2)}}
\end{align}
for occupied states
and 
\bal \label{nunu}
\prod_{j=1}^{d-n_1-m} \big(1-\zeta_j \big) \eq \prod_{k=1}^{n_0} \big(1-0 \big) \prod_{l=1}^{d-n_1 -n_0-m} \big(1-\zeta_l^{\sma{(1/2)}} \big) \lin
\eq \prod_{l=1}^{n_{1/2}-m} \big(1-\zeta_l^{\sma{(1/2)}} \big)
\end{align}
for unoccupied states.
We substitute Eq. (\ref{impi}) and (\ref{nunu}) into (\ref{suzy}) - up to exponential finite-size corrections, the set of nonzero eigenvalues in the particle-number sector $n_1 + m$ is
\bal
\bigg\{\lambda \bigg\}_{n_1 +m} \eq \bigg\{ \prod_{i =1}^{m} \zeta_i^{\sma{(1/2)}} \prod_{j=1}^{n_{1/2}-m} \big(1-\zeta_j^{\sma{(1/2)}} \big)\lin
&\;\;\;\bigg| \sum_{i} n^{\sma{(1/2)}}_i= m \bigg\}.
\end{align}
Once again, $i$ ($j$) labels occupied (unoccupied) entanglement states.
There are ${ n_{1/2} \choose m }$ combinations of single-particle occupation numbers that give nonzero values of $\lambda$ in the $n_1+m$ sector. 

We observe that the set of nonzero eigenvalues in the $n_1+m$ sector is independent of $n_1$. This result is applicable to both CI and QSHI. By inducing charge transport in either CI or QSHI through flux threading, the wavefunction only changes in the occupation numbers of edge states, hence by Thm. \ref{thm:1} only $n_1$ is affected, while $\zeta_i^{\sma{(1/2)}}$ are invariant. 

Suppose we began with a $N_{\sma{A}}= n_1 +m$ number sector. If $Q$ charges are transported from the right to the left edge due to flux threading, $n_1$ increases by $Q$ but the set of nonzero eigenvalues is invariant. The set of many-body entanglement eigenvalues in the $N_{\sma{A}}$ sector before flux threading is identical to the set of eigenvalues in the $N_{\sma{A}}+Q$ sector after flux threading. This statement is expressed mathematically in Eq. (\ref{specflowentspec}) and (\ref{eq:Z2flow}) in the contexts of the CI and QSHI respectively. The statement is true for all $N_{\sma{A}}$ so there is a global translation in particle-number space.    

\section{Proof that $T G(k_y) T^{-1} = G(-k_y)$ if the occupied subspace at $\pm k_y$ is closed under time-reversal} \label{cormatrixtrs}

We write the time-reveral operation as $T = V\, K$ with $V$ a unitary operator and $K$ complex conjugation. $T$ acts on spin-half single-particle states as $T^{2}=-I$. We define the time-reversal sewing matrix in the basis of occupied bands $F_{nm}(k_y) = \bra{u^n(-k_y)} T \ket{u^m(k_y)}; \;\;n,m = 1 \ldots N_{\sma{occ}}$
and another sewing matrix in the basis of all bands $H_{nm}(k_y) = \bra{u^n(-k_y)} T \ket{u^m(k_y)}; \;\;n,m = 1 \ldots N_{\sma{all}}$.
$F$ is unitary if the subspace of occupied bands is closed under time-reversal, i.e. $H$ decouples into two block diagonals for the occupied and unoccupied bands. Proof: Using the completeness property of the Bloch eigenvectors and the summation convention, we show that $H$ is unitary:
\bal
& {H}_{ji}^{\ast}(k_y) H_{jm}(k_y) \lin
\eq  u^j_m(-k_y)\,u^i_q(k_y) \,V^{\ast}_{pq} \;u^{j{\ast}}_r(-k_y)\,u^{m{\ast}}_s(k_y) \,V_{rs} = \delta_{im}
\end{align}
If $H$ decouples into two block diagonals for occupied and unoccupied bands, the unoccupied block $F$ must itself be unitary.
If $F$ is unitary, we derive two identities from the definition of $F$:
\bal
V_{ml}\, u^{i{\ast}}_m(-k_y) \eq F_{ij}(k_y)\,u^j_l(k_y) \lin
F_{il}(k_y)\, u^i_m(-k_y) \eq V_{mn} \, u_n^{l{\ast}}(k_y). 
\end{align}
From these two identities, $V = i\sigma_2 \otimes I_o = V^{\ast}$, and the definition of $G(k_y)$ in Eq. (\ref{cormatrix}),
\bal
&(T \,G(k_y))_{nl} = V_{nm}\,K \,u^{j{\ast}}_m(k_y)\,u^{j}_l(k_y) \lin
\eq V^{\ast}_{nm}\,u^{j}_m(k_y)\,u^{j{\ast}}_l(k_y) \,K  =  F^{\ast}_{ij}(k_y)\,u^{i{\ast}}_n(-k_y)\,u^{j}_l(k_y) \,K \lin
\eq u^{i{\ast}}_n(-k_y)\,u^{i}_m(-k_y) \,V^{\ast}_{ml} \,K = (G(-k_y) \,T)_{nl} \lin
&T \,G(k_y)\,T^{-1} = G(-k_y) \;\;\;\; \text{as\;advertised.}
\end{align}

\section{Defining $\Theta$ for even and odd $N_y$} \label{app:definetheta}

We assume periodic boundary conditions unless flux is inserted. If $N_y$ is even, $k_y=0$ and $k_y=\pi$ are allowed momenta for physical states, and we define $\Theta = n_{\sma{(s)}}(\Phi=0;k_y = 0) + n_{\sma{(s)}}(\Phi=0;k_y = \pi)$ as the sum of singly-degenerate entanglement eigenstates above the entanglement gap at $k_y=0$ and $\pi$.  

If $N_y$ is odd, $k_y=0$ is an allowed momentum for physical states but $k_y=\pi$ is not. However, upon adiabatic insertion of $\pi$ charge flux we now have antiperiodic boundary conditions, and the single-particle states at $k_y=\pi-\tfrac{\pi}{N_y}$ flow to $\pi$. Now $k_y=\pi$ is allowed but $k_y=0$ is not. Hence we define for odd $N_y$, $\Theta = n_{\sma{(s)}}(\Phi=0;k_y = 0) +  n_{\sma{(s)}}\big(\Phi=\tfrac{\Phi_0}{2};k_y = \pi\big)$ where the first term is evaluated with periodic boundary conditions and the second with antiperiodic boundary conditions.

\end{document}